\documentclass[12pt]{article}
\pdfoutput=1
\usepackage{subfigure}
\usepackage{amssymb,amsmath}
\usepackage{graphicx}
\usepackage{color}
\usepackage[colorlinks=true
,urlcolor=blue
,citecolor=blue
,linkcolor=blue
,pagecolor=blue
,linktocpage=true
,pdfproducer=medialab
]{hyperref}
\usepackage[a4paper,width=15.2cm]{geometry}

\makeatletter \renewcommand{\@dotsep}{10000} \makeatother

\newcommand{\beq}{\begin{equation}}
\newcommand{\eeq}{\end{equation}}
\newcommand{\bea}{\begin{eqnarray}}
\newcommand{\eea}{\end{eqnarray}}

\begin{document}



\begin{flushright}
OSU-HEP-14-07
\end{flushright}

\vspace*{0.4in}

\renewcommand{\thefootnote}{\fnsymbol{footnote}}

\begin{center}
\LARGE {\bf \large Flavor Symmetry Based MSSM (sMSSM):\\[-0.05in] Theoretical Models and Phenomenological Analysis
}
\end{center}

\vspace*{0.1in}

\begin{center}
{\large  K.S. Babu$^a$\footnote{Email:
babu@okstate.edu}, Ilia Gogoladze$^b$\footnote{Email: ilia@bartol.udel.edu}, Shabbar Raza$^c$\footnote{Email: shabbar@udel.edu} and Qaisar Shafi$^b$\footnote{Email: shafi@bartol.udel.edu}}

\end{center}

\begin{center}
\it $^a$Department of Physics, Oklahoma State University,
Stillwater, Oklahoma 74078, USA

\bigskip
\it $^b$Bartol Research Institute, Department of Physics and Astronomy,\\
 University of Delaware, Newark, DE 19716, USA

\bigskip
\it $^c$State Key Laboratory of Theoretical Physics and Kavli Institute of Theoretical Physics\\ China (KITPC), Institute of Theoretical Physics, Chinese Academy of Sciences,\\ Beijing 100190, P. R. China
\end{center}

\renewcommand{\thefootnote}{\arabic{footnote}}
\setcounter{footnote}{0}
\thispagestyle{empty}


\begin{abstract}

We present a class of supersymmetric models  in which symmetry considerations alone
dictate the form of the soft SUSY breaking Lagrangian. We develop a class of minimal models,
denoted as sMSSM -- for  flavor symmetry-based minimal supersymmetric standard model, which respect a grand
unified symmetry such as $SO(10)$ and a non-Abelian flavor symmetry $H$ which suppresses SUSY-induced flavor
violation.  Explicit examples are constructed with the
flavor symmetry being gauged  $SU(2)_H$ and $SO(3)_H$ with the three families transforming as
${\bf 2+1}$ and ${\bf 3}$ representations respectively.  A simple solution is found in the case of $SU(2)_H$
for suppressing the flavor violating $D$--terms based on an exchange symmetry.
Explicit models based on  $SO(3)_H$ without the $D$--term problem are developed.
In addition, models based on discrete non-Abelian flavor groups are presented which are automatically
free from $D$--term issues.  The permutation group $S_3$ with a ${\bf 2+1}$ family assignment,
as well as the tetrahedral group $A_4$ with a ${\bf 3}$ assignment are studied.  In all cases, a simple solution
to the SUSY CP problem is found, based on spontaneous CP violation leading to a complex quark mixing matrix.
We develop the phenomenology of the resulting sMSSM, which is controlled by seven soft SUSY
breaking parameters for both the  ${\bf 2+1}$  assignment and the ${\bf 3}$ assignment of fermion families. These models
are special cases of the phenomenological  MSSM (pMSSM), but with symmetry restrictions.  We { discuss} the
parameter space of sMSSM compatible with LHC searches, $B$ physics constraints and dark matter relic abundance.
Fine-tuning in these models is relatively mild, since all SUSY particles can have masses below about 3 TeV.

\end{abstract}

\newpage

\section{Introduction}

With the discovery of the Higgs boson, the search for supersymmetric particles which stabilize
its mass is expected to be a primary goal of the LHC experiments.  Extensive supersymmetric (SUSY)
particle searches by the ATLAS and CMS collaborations with data collected at 7 TeV and 8 TeV
center of mass energies have not revealed any trace of SUSY.  The resulting limits have constrained the masses of
the SUSY partners of the strongly interacting particles, the gluino and the squarks, to be greater than
about 1.4 TeV and 1.1 TeV respectively \cite{ATLAS,CMS}.  This in turn has constrained
specific versions of the minimal supersymmetric Standard Model (MSSM), such as the constrained
MSSM (cMSSM).  On the theoretical side, simplifying assumptions are often made for the form of the soft SUSY
breaking Lagrngian, as in the case of cMSSM inspired by minimal supergravity \cite{cMSSM}, which are not fully justified on symmetry principles.

In this paper we propose and develop a class of flavor symmetry--based minimal supersymmetric standard models, sMSSM for short, which on
one hand is predictive, but on the other hand is somewhat less constrained compared to the widely studied
cMSSM.  As we shall see, sMSSM has three more phenomenological parameters compared to cMSSM.
A large slice of parameter space of sMSSM remains unexplored experimentally, but much of this space is within reach of the LHC
with  the potential for discovering supersymmetry when it resumes its operation.

We define sMSSM as the minimal supersymmetric standard model with the most general soft SUSY breaking Lagrangian subject to the
following two symmetry requirements.
\begin{enumerate}
\item[(i)]  The parameters are
compatible with a grand unified symmetry such as $SO(10)$.
\item[(ii)] A non-Abelian flavor symmetry $H$ acts on the three families which help suppress excessive flavor changing neutral current processes (FCNC) mediated by the SUSY particles.
    \end{enumerate}
The motivations for each of these requirement are outlined below.

\subsection{Compatibility with a grand unified symmetry}

Grand unification \cite{GUT}, which is well motivated on several grounds, is one of the primary drivers of supersymmetric theories,
and is well supported by the observed merging of the three gauge couplings at an energy scale of $M_{\rm GUT} \simeq 2 \times 10^{16}$ GeV when extrapolated from their measured low energy values with low energy supersymmetry.  Among grand unified symmetry groups $SO(10)$ \cite{SO10} is particularly
attractive as it unifies all members of a family into a single 16-dimensional irreducible representation.  It predicts the existence of one
right-handed neutrino per family, and thus leads to small neutrino masses and mixings via the seesaw mechanism.  From the point of view
of soft SUSY breaking, compatibility with $SO(10)$ unified symmetry would  considerably reduce the number of soft masses of squarks and sleptons,
from fifteen in the case of Standard Model (SM) gauge symmetry (corresponding to the fifteen chiral multiplets of the SM) down to three. It would also provide a symmetry reason for the unification of
gaugino masses, reducing the relevant parameters  from three in the SM to one.
While sMSSM is defined by requiring compatibility with $SO(10)$, we do not attempt to construct here complete unified models based
on $SO(10)$ which would invariably bring in some model-dependence related to grand unified symmetry breaking.

If the grand unified symmetry
is chosen to be $SU(5)$, the soft SUSY breaking mass parameters will increase from three of $SO(10)$ to six, since each family of fermions and
their superpartners is assigned to ${\bf 10 + \overline{5}}$ under $SU(5)$. While the general sMSSM setup would allow for $SU(5)$ GUT,
here we focus on models compatible with $SO(10)$ GUT, as they are more restrictive and provide a natural understanding of neutrino masses.

\subsection{Non-Abelian flavor symmetry to suppress SUSY flavor violation}

The purpose of the non-Abelian flavor symmetry $H$, taken to be compatible with the GUT symmetry, is to suppress flavor changing neutral
currents without relying on ad hoc assumptions such as universality of squark (or slepton) masses at the GUT scale.  Without such a symmetry,
the fermion mass matrices will not align perfectly with the soft squared mass matrices of the corresponding SUSY partners.
This would lead to excessive flavor changing neutral currents mediated by SUSY particles \cite{susyfcnc}.  For example, box diagrams involving the gluino and squarks
would lead to $K^0-\overline{K^0}$ mixing, $B^0-\overline{B^0}$ mixing and $D^0-\overline{D^0}$ mixing, while loop diagram involving gauginos and
sleptons would lead to flavor changing decays such as $\mu \rightarrow e\gamma$.
From the $K^0-\overline{K^0}$ sector one obtains the following constraints \cite{ciuchini}:
\begin{equation}\label{limit}
\left|({\rm Re,~Im})(\delta^d_{LL})_{12} (\delta^d_{RR})_{12}\right|^{1/2} \leq (1.8 \cdot 10^{-3},~2.6 \cdot 10^{-4}) \left({\tilde{m} \over 1 ~{\rm TeV}}\right)~.
\end{equation}
Here $(\delta_{AB})_{ij} = (m^2_{AB})_{ij}/\tilde{m}^2$ is a flavor violating squark mass insertion parameter,
for $(A,B) = (L, R)$, with $\tilde{m}$ being the average mass of the relevant squarks $(\tilde{d}$
and $\tilde{s}$ in this case).  For this estimate the gluino mass was assumed to be equal to the average squark mass.
Now, the natural magnitude of the mixing parameters $(\delta^d_{LL})_{12}$ and $(\delta^d_{RR})_{12}$, in the
absence of additional symmetries, should be of order the Cabibbo angle, $\sim 0.2$.
The constraints from Eq. (\ref{limit}) strongly suggest that the squarks of the first two generations are highly
degenerate -- in the limit of exact degeneracy the mixing parameters $(\delta^d_{LL})_{12}$ etc would vanish.
The needed degeneracy is provided in sMSSM by a non-Abelian flavor
symmetry $H$ under which the $(\tilde{d},\,\tilde{s})$ fields belong to a common multiplet.
Analogous limits from $B_d^0-\overline{B_d^0}$ mixing are less severe, and are given by \cite{becirevic}:
\begin{equation}
\left|({\rm Re,~Im})(\delta^d_{LL})_{13} (\delta^d_{RR})_{13}\right|^{1/2} \leq (4.2 \cdot 10^{-2},~1.9 \cdot 10^{-2}) \left({\tilde{m} \over 1 ~{\rm TeV}}\right)~. \label{Bfcnc}
\end{equation}
Note that the natural value of this mixing parameter, in the absence of other symmetries, is $V_{ub} \sim
3 \times 10^{-3}$.  The constraints from Eq. (\ref{Bfcnc}) are well within limits.  $B_s-\overline{B}_s$ mixing
provides even weaker constraints.  This suggests that under the non-Abelian symmetry $H$, only the first two families need to form a common
multiplet in order to solve the SUSY flavor violation problem, while the third family could be a singlet.  We shall thus consider a ${\bf 2+1}$
assignment of fermion fields under $H$, as well as a ${\bf 3}$ assignment.  Both cases will lead to the same low energy phenomenology, as we shall see.

\subsection{The choice of non-Abelian flavor symmetry}

As for the choice of the flavor symmetry $H$, at first sight any symmetry group with doublet and/or triplet representations would
appear to suffice.  However, there are important restrictions on $H$.  Ideally, any symmetry should be a local gauge symmetry.
One is naturally led then to the gauge groups $SU(2)$, $SO(3)$ and $SU(3)$ which contain doublet and/or triplet representations.
Such gauge symmetries, however, potentially contain new sources of SUSY flavor violation, arising from the $D$-terms which split the masses of superparticles
within a given $H$--multiplet after SUSY breaking \cite{murayama}.  We propose a simple solution to this $D$-term problem in $SU(2)_H$ flavor symmetric models, based on an interchange symmetry acting on the Higgs doublet fields which break $SU(2)_H$.  In the case of $SO(3)_H$ models, it has
been noted in Ref. \cite{babubarr}
that the $D$-term problem can be controlled by breaking the symmetry with Higgs triplets which acquire real vacuum expectation
values.  We develop this class of models further, and show that with a simple discrete symmetry, the offending $D$-terms can be suppressed
completely.  We have not found a simple solution to the $D$-term problem if the flavor symmetry is gauged $SU(3)$.  It should be noted that
the low energy SUSY phenomenology is identical in the case of $SU(2)_H$ with a ${\bf 2+1}$ assignment of families, and in $SO(3)_H$ with a
${\bf 3}$ assignment.  This occurs because the ${\bf 3}$ assignment practically breaks down to a ${\bf 2+1}$ assignment
when  generating the observed top quark mass.

A variety of models based on flavor symmetries have been proposed to address the SUSY flavor problem in the literature
\cite{dine,local}.  In Ref. \cite{dine}, global $SU(2)$ and $SU(3)$ family symmetries were proposed. If the symmetry
is global, one has to deal with the Goldstone bosons associated with its spontaneous breaking.  Global symmetries are also susceptible to violations from quantum gravity. Clearly, local gauge symmetries are preferable, in which case the $D$-term problem should be addressed.
Some exceptions to the $D$-term problem have been noted in Ref. \cite{local}.  The sMSSM models proposed here are fully realistic models
without the $D$-term problem or Goldstone bosons, and { are} protected from flavor violation induced by quantum gravity effects.  They also
shed some light on the fermion mass hierarchy puzzle.

An interesting alternative to local gauge symmetry is to identify $H$ with a non-Abelian discrete symmetry with either ${\bf 2+1}$
representations or ${\bf 3}$ representations \cite{seiberg,babu}.  With such a  choice of $H$ there is no $D$-term problem at all.
Such non--Abelian discrete symmetries have found application in understanding the various puzzles associated with the quark and lepton
masses and mixing angles with or without supersymmetry \cite{other}.  More
recently, such symmetries have been used to understand the near tri--bimaximal neutrino mixing pattern \cite{othernu}.  It would be desirable
to find a symmetry that sheds light on the fermion mass and mixing puzzle, and at the same time solves
the SUSY flavor problem.  Here we present explicit examples with the permutation group $S_3$ and the tetrahedral group $A_4$.
The group $S_3$ admits ${\bf 2+1}$ assignment of fermion families, while $A_4$ admits a ${\bf 3}$ assignment.  We focus on these
groups as they are the simplest discrete groups with doublet and triplet representations. A nontrivial task in building such models
is to make sure that they do lead to realistic fermion masses and mixings, which would require a consistent symmetry breaking mechanism.
In all our models we address this issue.   As in the case of $SU(2)_H$ and
$SO(3)_H$, these discrete groups lead to the same low energy SUSY phenomenology parametrized by sMSSM.  Other non-Abelian discrete groups such
as $D_n$ and $Q_{2n}$ for integer values of $n$ would yield similar results.

\subsection{The SUSY CP problem and a solution}

Supersymmetric models { face} another problem, related to CP violation.  There are new sources of CP violating phases,
arising from the soft SUSY breaking sector. The new phases should be less than about ($10^{-1}-10^{-2}$), depending on the squark and
slepton masses and the value of $\tan\beta$, to be consistent with electric dipole moment limits on the neutron and the electron \cite{edm}.
The Kobayashi-Maskawa (KM) phase in the quark sector,
on the other hand, is of order unity.  This disparity defines the SUSY CP problem.  In our models, we require spontaneous
CP violation, in which case the soft parameters will all be real. In some cases, a phase alignment is observed in the Yukawa matrix
and the trilinear $A$-term matrix, which suppresses the CP phases sufficiently \cite{babu}.  Even in  cases where the phases in these two
matrices do not align, the SUSY phase problem is significantly ameliorated.  An order one phase is induced in the quark mixing matrix, so that
the success of the CKM paradigm is preserved.  A simple way of introducing spontaneously induced phases into the quark mixing matrix
is suggested, making use of a singlet scalar field which acquires a complex vacuum expectation value.
The phenomenology of sMSSM assumes all soft parameters to be real, as in the case of most
cMSSM analyses.

\subsection{Phenomenological analysis}

We perform a numerical analysis of the allowed SUSY parameter space of sMSSM.
This parameter set is slightly larger (by three) than the four (five if sgn($\mu$) is counted)
parameters used in cMSSM, but still quite restrictive.
Specifically, sMSSM is defined by the following parameter set:
\begin{equation}
\{m_{{0}_{(1,2)}},\, m_{{0}_{(3)}},\, M_{1/2},\, A_0,\,\tan\beta,\, |\mu|, \, m_{A},\, {\rm sgn}(\mu)\}~.
\label{para}
\end{equation}
Here $m_{{0}_{(1,2)}}$ is the common mass of the first two family sfermions, while $m_{{0}_{(3)}}$ is that of the
third family sfermions.  $m_A$ is the mass of the pseudoscalar Higgs boson, and $M_{1/2}$ is the universal gaugino mass.
Recall that the cMSSM  is defined by the parameter set $\{m_{0},\, M_{1/2},\, A_0,\,\tan\beta,\, {\rm sgn}(\mu)\}$,
which has three fewer parameters than the set of sMSSM.
We delineate the parameter space of sMSSM that is compatible with radiative electroweak symmetry breaking, direct
LHC search limits, $B$ physics constraints and relic abundance of dark matter.
Our analysis shows that all SUSY particle masses can be below about 3 TeV, resulting in relatively mild fine-tuning.
Continued search for SUSY at the LHC should reveal almost every superpartner in sMSSM.

sMSSM is a special case of phenomenological MSSM (pMSSM) \cite{pMSSM} which has been widely studied, but with additional symmetry restrictions.  Specifically, the subset of pMSSM spectrum that is consistent with a grand unified symmetry such as $SO(10)$ will resemble the spectrum of sMSSM.
The number of parameters in sMSSM (seven) is much smaller than the corresponding number in pMSSM (nineteen), which makes
exploration of the full parameter space more manageable in sMSSM.  Minimal SUSY models allowing for non-universal Higgs masses
with $m_{H_u}^2 \neq m_{H_d}^2$ have been studied under the acronym NUHM2 (non-universal Higgs masses -- 2) \cite{NUHM2}.
These models would reduce to sMSSM, if the third family soft scalar mass at the GUT scale is assumed to be different from
that of the first two families.

The rest of the paper is organized as follows.
In Section 2 we motivate and develop the sMSSM models.  There we present UV complete theories based on $SU(2)_H$ and $SO(3)_H$ local
gauge symmetries.  A new solution to the $D$-term problem is proposed for the $SU(2)_H$ models.  Models based on $SO(3)_H$ are
developed and shown to be consistent with flavor changing constraints.  Two models based on discrete non-Abelian symmetries
$S_3$ and $A_4$ are also presented.  In all these models, a solution to the SUSY phase problem is noted, based on spontaneous CP violation
leading to a complex CKM matrix.  In Section 3 we explore the parameter space of sMSSM models, requiring consistency with direct
search limits on SUSY particles, radiative electroweak symmetry breaking, $B$ physics and relic abundance of cold dark matter.  In Section 4 we have our concluding remarks.

\section{Explicit Models Leading to sMSSM}

The primary goal of sMSSM is to understand in a controlled fashion, based on underlying symmetries, the breaking of supersymmetry.
The Lagrangian of these models is the most general compatible with specified symmetries.  Such a setup is guaranteed to be stable against potential Planck scale corrections in the K\"ahler potential as well as the superpotential, provided that the symmetries have a gauge origin.

sMSSM requires two symmetries, as noted in the introduction.  The first, grand unified symmetry, is very powerful in restricting the number of free parameters.  For a variety of reasons, $SO(10)$ \cite{SO10} appears to us to be the GUT of choice, owing to the essential requirement of the right-handed neutrino (to complete the spinor multiplet of $SO(10)$)  needed for the seesaw mechanism and possibly for leptogenesis.  Under $SO(10)$, the  quarks and leptons of each family and their superpartners
transform as ${\bf 16}$--plets of which there are three copies.  This GUT symmetry imposes the restriction that the three
gaugino masses should be unified, so long as $SO(10)$ gauge singlets drive supersymmetry breaking, which is what we assume.  Similarly, the soft SUSY breaking masses of all members of a ${\bf 16}$--plet must be degenerate at the GUT scale, owing to $SO(10)$. Note, however, that the soft masses of the Higgs fields $m^2_{H_u}$ and $m^2_{H_d}$ are not required to be the same as that of the ${\bf 16}$--plets, as $H_u$ and $H_d$ belong to ${\bf 10}$--dimensional representations of $SO(10)$.  Most realistic $SO(10)$ models also utilize ${\bf 16}$ and ${\bf \overline{16}}$--plet Higgs fields for symmetry breaking \cite{bpt}, which also contain $H_u$ and $H_d$--like components so that the MSSM fields $H_u$ and $H_d$ are partially in ${\bf 10}$ and partially in ${\bf 16}+ {\bf \overline{16}}$. Thus the soft SUSY breaking masses of $H_u$ and $H_d$ are not required to be the same, as there is
no reason for the soft masses of ${\bf 10}$--plet and ${\bf 16}+{\bf \overline{16}}$--plets to be the same, and the admixtures of these fields in the $H_u$ and $H_d$ sectors are in general unequal.  Note that these models do not require $\tan\beta \simeq m_t/m_b$, owing to the admixture of
${\bf 16}$ in the light field $H_d$ of MSSM \cite{bpt}.

While $SO(10)$ symmetry already constrains the SUSY breaking parameters significantly, it is not sufficient to suppress potentially large FCNC processes mediated by the SUSY particles.  sMSSM overcomes this problem with a flavor symmetry $H$.  This flavor symmetry unifies the three ${\bf 16}$--plets into either a ${\bf 2+1}$ pattern or a ${\bf 3}$ pattern.
Such a unification of families would make the soft masses of the first two families degenerate in the case of a ${\bf 2+1}$
pattern, and all three families degenerate in the case of a triplet pattern.  It turns out that both cases would result in seven effective SUSY breaking parameters with the FCNC processes sufficiently under control.

The symmetry group $H$ must have doublet or triplet representations.  If $H$ is a local gauge symmetry, the choices are $SU(2)_H$,
$SO(3)_H$ and $SU(3)_H$.  In $SU(2)_H$ and $SO(3)_H$ models the triangle gauge anomalies automatically vanish, but not in $SU(3)_H$ models.
As we show below, there are simple solutions to the $D$-term problem in $SU(2)_H$ and $SO(3)_H$ models, but we have been unable to extend
this to the case of $SU(3)_H$.  Thus we focus on the former two gauge groups.  We also develop models based on non-Abelian discrete symmetries.
The natural choices are $S_3$, the permutation group of three letters,
which admits a ${\bf 2+1}$ family assignment, and the tetrahedral group $A_4$, which
admits a ${\bf 3}$ family assignment.  There are no $D$-terms in these cases, and thus no $D$-term problem.
If such symmetries are discrete remnants of a true gauge symmetry quantum gravity corrections will not violate the symmetry.  It should be noted that string compctification often provides discrete non-Abelian symmetries such as $S_3$, $Q_4$ and $A_4$, which are of the type sMSSM models would need.

\subsection{sMSSM from \boldmath{$SU(2)_H$} flavor symmetry}

Here we propose and develop a gauge model based on $SU(2)_H$ flavor symmetry acting on the three families of ${\bf 16}$.
(We shall use a compact notation using $SO(10)$ language, although such an underlying GUT is not necessary for the discussions to be valid.)
The fermions and their superpartners are assigned to a ${\bf 2+1}$ representation under $SU(2)_H$.  That is, $({\bf 16_1}, \, {\bf 16}_2)$ form
a doublet under $SU(2)_H$, while ${\bf 16_3}$ is a singlet.  More explicitly, the assignment of fermion families under $SU(2)_H$ is as follows:
\begin{eqnarray}
&{\bf 2}:& ~~\vec{Q} = \left(\begin{matrix}q_1 \\ q_2  \end{matrix}\right);~~ \vec{L} = \left(\begin{matrix}\ell_1 \\ \ell_2  \end{matrix}\right);~~
\vec{u^c} = \left(\begin{matrix}u_1^c \\ u_2^c  \end{matrix}\right);~~ \vec{d^c} = \left(\begin{matrix}d_1^c \\ d_2^c  \end{matrix}\right);~~
\vec{e^c} = \left(\begin{matrix}e_1^c \\ e_2^c  \end{matrix}\right);~~ \vec{\nu^c} = \left(\begin{matrix}\nu_1^c \\ \nu_2^c  \end{matrix}\right);
\nonumber\\[0.15in]
&{\bf 1}:& ~~ q_3;~~ \ell_3;~~ u_3^c; ~~ d_3^c;~~ e_3^c;~~ \nu_3^c.
\end{eqnarray}
The soft SUSY breaking Lagrangian must respect this symmetry, which would imply degeneracy of $(\tilde{q}_1,\,\tilde{q}_2)$ masses, for example.

The $SU(2)_H$ symmetry must be broken spontaneously in order to generate realistic quark and lepton masses.
The symmetry breaking sector relevant at high energies, where we assume the flavor symmetry breaks spontaneously, consists of a pair of  $SU(2)_H$ doublets denoted as $\phi$ and $\overline{\phi}$ which are singlets of the Standard Model
and $SO(10)$. The superpotential involving these fields is given by
\begin{equation}
W_{\rm sym} = \mu_\phi\, \phi \,\overline{\phi} + \kappa \,(\phi\, \overline{\phi})^2~.
\end{equation}
Here $\kappa$ is a parameter with inverse dimension of mass, which can originate either from quantum gravity corrections proportional inversely to
the Planck mass, or by integrating out a $SO(10) \times SU(2)_H$ gauge singlet field.  It is possible to keep such a singlet field in the spectrum,
which does not change our conclusions.  Denoting the doublet fields as
\begin{eqnarray}
\phi = \left(\begin{matrix} \phi_1 \cr \phi_2 \end{matrix} \right),~~\overline{\phi} = \left(\begin{matrix} \overline{\phi}_2 \cr \overline{\phi}_1 \end{matrix} \right),
\end{eqnarray}
the scalar potential can be written down as $V= V_F + V_D + V_{\rm soft}$ where
\begin{eqnarray}
V_F &=& \left(|\phi_1|^2+|\phi_2|^2+|\overline{\phi}_1|^2 + |\overline{\phi}_2|^2\right)\,\left|\mu_\phi - 2 \kappa(\phi_1 \overline{\phi}_1 - \phi_2 \overline{\phi}_2) \right|^2,
\label{potF} \\
V_D &=& \frac{g_H^2}{8}\left[ |\phi_1^* \phi_2 + \phi_2^* \phi_1 + \overline{\phi}_2^*\, \overline{\phi}_1 + \overline{\phi}_1^* \,
\overline{\phi}_2 |^2 + | \phi_1^* \phi_2 - \phi_2^* \phi_1 + \overline{\phi}_2^*\, \overline{\phi}_1 - \overline{\phi}_1^* \,
\overline{\phi}_2 |^2 \right. \nonumber \\
&~& \left. + | \phi_1^* \phi_1 - \phi_2^* \phi_2 + \overline{\phi}_2^*\, \overline{\phi}_2 - \overline{\phi}_1^* \,
\overline{\phi}_1 |^2 \right], \label{potD} \\
V_{\rm soft} &=& m_\phi^2(|\phi_1|^2+ \phi_2|^2) + m_{\overline{\phi}}^2(|\overline{\phi}_1|^2+ (|\overline{\phi}_2|^2) \nonumber \\
&+& \{ - B \mu_\phi(\phi_1 \overline{\phi}_1 - \phi_2 \overline{\phi}_2) + C\kappa(\phi_1 \overline{\phi}_1 - \phi_2 \overline{\phi}_2)^2 + {\rm h.c.} \}
\label{potsoft}
\end{eqnarray}
This potential admits a vacuum solution given by
\begin{eqnarray}
\left\langle \phi \right \rangle =
 \left(\begin{matrix}0 \\ u  \end{matrix}\right);~~\left\langle \overline{\phi} \right\rangle = \left(\begin{matrix}\overline{u} \\ 0  \end{matrix}\right)~
\end{eqnarray}
with $u$ and $\overline{u}$ determined by the two complex extremum conditions,
\begin{eqnarray}
0 &=& u^*\left|\mu_\phi + 2 \kappa u \overline{u}\right|^2 + 2 \kappa \overline{u}(|u|^2+|\overline{u}|^2)(\mu_\phi + 2 \kappa u \overline{u})^*
+\frac{g_H^2}{4} u^*(|u|^2-|\overline{u}|^2) \nonumber \\
&+& m_\phi^2 u^* + B \mu_\phi \overline{u}+2 C \kappa u \overline{u}^2, \nonumber \\
0 &=& \overline{u}^*\left|\mu_\phi + 2 \kappa u \overline{u}\right|^2 + 2 \kappa u(|u|^2+|\overline{u}|^2)(\mu_\phi + 2 \kappa u \overline{u})^*
-\frac{g_H^2}{4} \overline{u}^*(|u|^2-|\overline{u}|^2) \nonumber \\
&+& m_{\overline{\phi}}^2 \overline{u}^* + B \mu_\phi u+2 C \kappa  \overline{u} u^2.
\label{min}
\end{eqnarray}

In the supersymmetric limit, i.e., when $\{m_\phi^2,\, m^2_{\overline{\phi}},\, B,\, C\}$ are set to zero in Eqs. (\ref{min}), we have
\begin{equation}
|u| = |\overline{u}|;~~~~\mu_\phi + 2 \kappa u \overline{u} = 0~.
\end{equation}
Including SUSY breaking terms, assumed to be much smaller in magnitude compared to $|u|$, we obtain from Eq. (\ref{min}) a relation
\begin{equation}
|u|^2 - |\overline{u}|^2 = \frac{2 (m^2_{\overline{\phi}} - m_\phi^2)}{g_H^2}
\left(1+ {\cal O}\left\{ \frac{|B \mu_\phi|}{|u|^2},\,\frac{|C \mu_\phi|}{|u|^2}\right\}\right)~.
\label{split}
\end{equation}
This non-vanishing $D$-term causes a splitting in the masses of the first two family squarks, which is given by
(using $\{|B\mu_\phi|,\,|C\mu_\phi| \} \ll |u|^2$)
\begin{equation}
{\cal L}_D = \frac{m_\phi^2 - m^2_{\overline{\phi}}}{2} \left(|\tilde{q}_1|^2 - |\tilde{q}_2|^2\right)~.
\label{D-term}
\end{equation}
Note that the $SU(2)_H$ gauge coupling $g_H$ has disappeared in Eq. (\ref{D-term}), so that even by choosing small
values of $g_H$ the squark mass splitting will be of the same order as the squark mass itself.  This is the $D$-term
problem of gauged flavor symmetry, as this splitting would induce excessive flavor violation in conflict with the limits shown in Eq. (\ref{limit}).

\subsubsection{Solution to the \boldmath{$D$}-term problem}

Here we propose a simple solution to the $D$-term problem of $SU(2)_H$ models.  If an interchange symmetry $\phi \leftrightarrow
\overline{\phi}$ is imposed, then $m^2_\phi = m^2_{\overline{\phi}}$ in Eq. (\ref{potsoft}).
Eq. (\ref{split}) would then imply that $|u|^2 = |\overline{u}|^2$,
and there is no $D$-term splitting problem.  Such an interchange symmetry is quite natural.  In fact, the model under discussion is
a gauged $SU(2)_H$ model with two flavors ($\phi$ and $\overline{\phi}$).  The anomaly free global symmetry of the model is
$SU(2)$, which rotates the two doublets.  The interchange symmetry $\phi \leftrightarrow
\overline{\phi}$ is a subgroup of this anomaly free global $SU(2)$.  No quantum corrections will spoil this symmetry.
Note that the Higgs potential of Eqs. (\ref{potF})-(\ref{potsoft}) and the gauge interactions respect this interchange symmetry.
As we shall show below, realistic fermion masses can be generated consistent with this interchange symmetry.

It is worthwhile to note that such an interchange symmetry is possible only because the doublet representation of
$SU(2)_H$ is pseudoreal.  $\phi$ and $\overline{\phi}$ have identical gauge properties in $SU(2)_H$.  If the family
gauge symmetry were $SU(3)_H$ with the Higgs fields transforming as ${\bf 3+ \overline{3}}$, such an interchange symmetry
will not work in any simple way, as the ${\bf 3}$ of $SU(3)$ is distinct from the ${\bf \overline{3}}$.  This is why we
are unable to generalize our solution to $SU(3)_H$ $D$-terms. If the flavor symmetry is $SO(3)$ broken by Higgs triplets,
such an interchange symmetry is a possible solution to the $D$-term problem, although in the next subsection we develop
an alternative solution for the case of $SO(3)_H$.

\subsubsection{Realistic fermion masses}

We can write down the superpotential relevant for fermion mass generation that is consistent with $SU(2)_H$ gauge symmetry
and the interchange symmetry $\phi \leftrightarrow \overline{\phi}$.  The fermion fields are all invariant under the
interchange symmetry.  In the notation of $SO(10)$ the superpotential takes the form
\begin{equation}
W_{\rm Yuk} = {\bf 16}_3 {\bf 16}_3 {\bf 10}_H + {\bf 16}_i {\bf 16}_3 {\bf 10}_H \left(\frac{\phi_j +\overline{\phi}_j}{M_*}\right) \epsilon^{ij} +
{\bf 16}_i {\bf 16}_j \epsilon^{ij} {\bf 10}_H \left(\frac{{\bf 45}_H}{M_*}\right) + ...
\label{Yuk1}
\end{equation}
Here $...$ stands for higher order terms suppressed by more powers of $M_*$, which is presumable the Planck scale, much
larger than $\left\langle {\bf 45}_H \right \rangle \sim M_{\rm GUT}$ and $|u|$.  As emphasized earlier, it is not required that
the  $SO(10)$ GUT symmetry is employed.  If it is indeed used, the coupling ${\bf 16}_i {\bf 16}_j \epsilon^{ij} {\bf 10}_H$
will not be allowed owing to the clash between the symmetric nature of this terms under $SO(10)$ and antisymmetry under $SU(2)_H$.
However, the combination
shown in Eq. (\ref{Yuk1}) with an additional ${\bf 45}_H$ would be permitted, owing to its $SO(10)$-antisymmetric component. Note that superpotential in Eq. (\ref{Yuk1}) give a desirable framework  for $t-b-\tau$ Yukawa coupling unification as well \cite{yukawaUn, Ajaib:2013zha}.

Now, the minimization of the potential for $\phi+\overline{\phi}$ with the interchange symmetry would lead to the vacuum solution
\begin{equation}
\left\langle \phi + \overline{\phi} \right\rangle = u \left(\begin{matrix} e^{i \alpha} \\ 1 \end{matrix} \right)~.
\end{equation}
A common rotation in the 1-2 family space can be used to bring this to the form proportional to $(0,\,1)^T$.  Such a rotation
would leave the 1-2 sector of the fermion mass matrix invariant, as it is proportional to the second Pauli matrix $\tau_2$.
(Recall that $U \tau_2 U^T = \tau_2$ for arbitrary unitary matrix $U$.) Thus one arrives
at  fermion mass matrices of the form
\begin{eqnarray}
M_f = \left( \begin{matrix} 0 & c & 0 \\ -c & 0 & b \\ 0 & b' & a \end{matrix}
\label{Mf1}
\right)_f
\end{eqnarray}
for $f=u,\,d,\,\ell,\,\nu^D$. In the up-quark mass matrix, the entry $a$ is of order the top quark mass, arising from
the unsuppressed first operator of Eq. (\ref{Yuk1}).  The entries $b,\,b'$ are smaller, suppressed by a factor
$|u|/M_*$.  Finally, the entry $c$ is suppressed by $M_{\rm GUT}/M_*$, which can be much smaller.  Thus we see that
the model provides a qualitative understanding of the fermion mass hierarchy.

Mass matrices of the form of Eq. (\ref{Mf1}) give an excellent fit to the observed masses and mixing angles, as shown
in Ref. \cite{babu}.  The inter-family mixing angles in each sector are small, so in diagonalzing these mass matrices,
excessive flavor violating couplings in the squark and slepton sectors are not induced.  The leading form of the squark
mass matrix in the original basis is
\begin{eqnarray}
M^2_{\tilde{f}} = \left(\begin{matrix} m_1^2 & ~ & ~ \\
~ & m_1^2 & ~ \\ ~ & ~ & m_3^2   \end{matrix} \right)~.
\end{eqnarray}
When the fermions are brought into their mass eigenstates, this form will be essentially maintained, with very small
flavor violating couplings which are consistent with the limits given in Eq. (\ref{limit}).

The leading higher dimensional operator that can lead to flavor violation arises from the K\"ahler
potential, and is of the type
\begin{equation}
{\cal L} \supset \int d^4\theta\left\{ (\vec{Q}^\dagger \phi)(\overline{\phi}^\dagger \vec{Q})  \frac{|Z|^2}{M_{\rm Pl}^4} + \phi \rightarrow \overline{\phi}\right\}~.
\label{kahler1}
\end{equation}
When a nonzero $F$-component of the spurion field $Z$ that breaks supersymmetry is inserted in Eq. (\ref{kahler1}), an off-diagonal
entry in the squark mass matrix will be induced, with
a magnitude of order $m_{\rm SUSY}^2 |u|^2/M_{\rm Pl}^2 \sim 10^{-4} \,m_{\rm SUSY}^2$.  We see that such flavor violations are
sufficiently suppressed, and consistent with the limits of Eq. (\ref{limit}).  Thus the SUSY flavor
violation problem is resolved in the model.

It should be noted that symmetry considerations alone would allow the trilinear $A$-terms of the SUSY breaking Lagrangian to
be non-proportional to the corresponding Yukawa couplings.  However, the $A$-terms would have the same chirality suppression
as the fermion Yukawa couplings, and thus would not cause excessive SUSY flavor violation.  Processes such as $\mu \rightarrow
e\gamma$ would be expected near the current experimental limit nevertheless, owing to chirally suppressed contributions from
non-proportional $A$-terms \cite{babu}.  For collider phenomenology it is only the third family $A$-terms that are relevant.
$A_0$ in Eq. (\ref{para}) is defined as $A_0 \equiv A_t^0 = A_b^0 = A^\tau_0$.  In general these three parameters need not be
equal at the GUT scale, however, any difference will play a role only for large values of $\tan\beta$.  In our phenomenological
analysis we shall assume equality of the third family $A$-terms.

\subsubsection{Solution to the SUSY CP problem}

We note that the SUSY CP problem can be resolved in our setup by requiring that CP be a spontaneously broken symmetry.
Nevertheless, the quark mixing matrix will be complex, and the success of the CKM CP violation will be maintained.

In order to generate spontaneous CP violation, we note that a complete singlet superfield $X$ can have a superpotential
\begin{equation}
W(X) = a X + \frac{b}{2} X^2 + \frac{c}{3} X^3~.
\end{equation}
Demanding $F_X = 0$ gives
\begin{equation}
\left\langle X \right \rangle = \frac{1}{2c}\left(-b \pm \sqrt{b^2 - 4 ac}\right)~.
\end{equation}
Suppose that CP is a good symmetry.  In this case the parameters $(a,\,b,\,c)$ are all real. However, if $b^2-4ac < 0$
the vacuum expectation value $\left\langle X \right \rangle$ will be complex.  Now $X$ being a gauge singlet couples
to other fields through renormalizable couplings such as $\phi \overline{\phi} X$ and  ${\bf 45}_H^2 X$.  When the
VEV of $X$ is inserted in these couplings, the effective mass terms for these fields would become complex
and minimization with respect to the other fields would generate complex phases in their VEVs as well. These complex
VEVs enter into the mass matrix of Eq. (\ref{Mf1}), and generate CKM CP violation.  Note that the soft SUSY breaking
parameters are all real, owing to  CP symmetry.  The mass matrix of Eq. (\ref{Mf1}) has an interesting feature.
Its phases can be factored out.  This means that the trilinear $A$-term matrices, with a similar phase structure that
can also be factored, would become real in a basis where the fermion masses are real and diagonal \cite{babu}.  Such a phase alignment would
solve the SUSY CP problem quite naturally.

\subsection{sMSSM from \boldmath{$SO(3)_H$} flavor symmetry}

In this section we present a model based on gauged $SO(3)_H$ flavor symmetry without the $D$-term problem.  The three families
of quarks and leptons transform as a ${\bf 3}$ under $SO(3)_H$.  Thus the fermions belong to a $({\bf 16,\,3})$
representation under $SO(10) \times SO(3)_H$.  Such an assignment is free from gauge anomalies.
Although this assignment would require that the soft SUSY breaking masses of all scalars
are the same at the GUT scale, we shall see that effectively the assignment  splits into a reducible {\bf 2+1} representation of a residual
$SO(2)_H$ symmetry, and thus would lead to the spectrum of sMSSM.  The top quark mass essentially breaks the flavor symmetry down to an
$SO(2)_H$ subgroup.

\subsubsection{Symmetry breaking and a solution to the \boldmath{$D$}-term problem}

$SO(3)_H$ symmetry is assumed to be broken spontaneously at a high energy scale, of order $M_{\rm GUT}$, by
 SM and $SO(10)$ singlet fields $\vec{A}_i$ belonging to ${\bf 3}$ of $SO(3)_H$.  A minimum of three such
 triplets will be used so that  $SO(3)_H$ breaks completely, and realistic and hierarchical fermion masses
 are generated.  We denote these triplets as $\vec{A},\, \vec{B},\,$ and $\vec{C}$ in the familiar vector notation
 which is applicable to $SO(3)$.  The most general superpotential involving these fields (after diagonalizing the
 bilinear terms) is
 \begin{equation}
 W = \frac{\mu_A}{2}\, \vec{A} \cdot \vec{A} + \frac{\mu_B}{2}\, \vec{B} \cdot \vec{B} +\frac{\mu_C}{2}\, \vec{C} \cdot \vec{C} + \lambda\,
 \vec{A} \times \vec{B} \cdot \vec{C}~.
 \label{W2}
 \end{equation}
 The scalar potential contains $F$-terms derived from Eq. (\ref{W2}), as well as $SO(3)_H$ $D$-terms and soft SUSY breaking terms.
 The $D$-term potential is given by
 \begin{equation}
V_D =  \frac{g_H^2}{2} \left|\sum_n \vec{\phi}_n^* \times \vec{\phi}_n + \sum_i \vec{A}_i^* \times \vec{A}_i \right|^2
 \end{equation}
 where the sum over $n$ takes into account all scalar of MSSM, and the sum over $i$ includes the symmetry breaking fields
 $\vec{A},\,\vec{B},\,\vec{C}$.  Note that $SO(3)_H$ allows for a description of $D$-terms in terms of vector cross products.
 Based on this fact, a solution to the $D$-term problem was suggested in Ref. \cite{babu}, which utilizes the smallness of
 CP violating phases in SUSY. Indeed, the term $\vec{A}_i^* \times \vec{A}_i = 2 \,i\, ({\rm Re}\vec{A}_i )\times ({\rm Im}\vec{A}_i)$
 would vanish in the limit of real VEVs for $\vec{A}_i$.  Here we suggest an alternative solution which would suppress the
 $D$-terms completely, even with complex VEVs for the $\vec{A}_i$ fields.

 The soft SUSY breaking Lagrangian of the model takes the form
 \begin{eqnarray}
 V_{\rm soft} &=& m_A^2 \vec{A}^\dagger \cdot \vec{A}+ m_B^2 \vec{B}^\dagger \cdot \vec{B}+ m_C^2 \vec{C}^\dagger \cdot \vec{C} + \left\{\lambda A_\lambda \vec{A} \times
 \vec{B} \cdot \vec{C} \right.\nonumber \\
 &~& \left. + \frac{B_A \mu_A}{2} \vec{A} \cdot \vec{A} + \frac{B_B \mu_B}{2} \vec{B} \cdot \vec{B} + \frac{B_C \mu_C}{2} \vec{C} \cdot \vec{C} + h.c.  \right\}
 \end{eqnarray}
 Note that the superpotential of Eq. (\ref{W2}) respects a discrete $Z_2 \times Z_2$ symmetry under which the fields transform as
 $\vec{A}: (-+)$, $\vec{B}: (+-)$ and $\vec{C}: (--)$. We have adopted this symmetry for the full Lagrangian so that the soft SUSY breaking terms do
 not contain cross terms of the type $\vec{A}^\dagger \cdot \vec{B}$.  This symmetry will be sufficient to make the $D$-terms of $SO(3)_H$ vanish.

 The potential of the model, including soft SUSY breaking terms, admits a vacuum structure of the form
 \begin{eqnarray}
 \left\langle \vec{A} \right \rangle = \left(\begin{matrix} 0 \\ 0 \\ u_3 \end{matrix}  \right),~~~~
 \left\langle \vec{B} \right \rangle = \left(\begin{matrix} 0 \\ u_2 \\ 0 \end{matrix}  \right),~~~~
 \left\langle \vec{C} \right \rangle = \left(\begin{matrix} u_1 \\ 0 \\ 0 \end{matrix}  \right)~.
 \label{vev}
 \end{eqnarray}
 The VEVs $u_i$ are given in the SUSY limit as
 \begin{equation}
 u_1 = \frac{\sqrt{\mu_A \mu_B}}{\lambda},~~~ u_2 = \frac{\sqrt{\mu_A \mu_C}}{\lambda},~~~ u_3 = \frac{\sqrt{\mu_B \mu_C}}{\lambda}~.
 \label{ui}
 \end{equation}
 A hierarchy $u_1 \ll u_2 \ll u_3$ can then be realized, which will be useful for understanding the fermion masses.
 Essentially, $u_3$ would define the third family direction in family space, while $u_2$ and $u_1$ would define the
 second and first family directions.

 The $SO(3)_H$ $D$-terms all vanish for the configuration of Eq. (\ref{vev}), since Re($\vec{A}$) and Im($\vec{A}$) are
 parallel vectors, and similarly for $\vec{B}$ and $\vec{C}$.  The $Z_2 \times Z_2$ symmetry plays a crucial role in realizing
 this structure.  This symmetry plays another important role in suppressing  SUSY flavor violation to the required order.
 Without this symmetry, a Lagrangian term of the type
 \begin{equation}
{\cal L} \supset \int d^4\theta \, \vec{Q}^\dagger_i\, \vec{Q}_j \, \vec{A}_k\, \epsilon^{ijk} \frac{|Z|^2}{M_{\rm Pl}^3}
\label{kahler2}
\end{equation}
would be allowed.  This would yield an off-diagonal squark mixing of order $m_{\rm SUSY}^2 |u_3|/M_{\rm Pl}$ $\sim
10^{-2} \,m^2_{\rm SUSY}$, in violation of the limit quoted in Eq. (\ref{limit}).  With the $Z_2 \times Z_2$ symmetry
which acts trivially on the MSSM fields, the term in Eq. (\ref{kahler2}) is forbidden, and the leading K\"ahler potential
correction to SUSY FCNC arises from the Lagrangian
 \begin{equation}
{\cal L} \supset \int d^4\theta  (\vec{Q}^\dagger \cdot \vec{B}) (\vec{B}^\dagger \cdot \vec{Q}) \frac{|Z|^2}{M_{\rm Pl}^4}
\label{kahler3}
\end{equation}
which would lead to consistent off-diagonal mixing of order $m_{\rm SUSY}^2 |u_2|^2/M_{\rm Pl}^2 \sim 10^{-4}\,m_{\rm SUSY}^2$.

The VEV structure of Eq. (\ref{vev}) is quite stable against higher dimensional operators in the superpotential.  The leading
corrections to $W$ take the form
\begin{eqnarray}
W' &=& \kappa_A (\vec{A} \cdot \vec{A})^2 + \kappa_B (\vec{B} \cdot \vec{B})^2 + \kappa_C (\vec{C} \cdot \vec{C})^2 \nonumber \\
&+& \kappa_{AB} (\vec{A} \cdot \vec{B})^2 + \kappa_{AC} (\vec{A} \cdot \vec{C})^2 + \kappa_{BC} (\vec{B} \cdot \vec{C})^2 ~.
\end{eqnarray}
Here $\kappa_i$ and $\kappa_{ij}$ have inverse dimensions of mass.  It is easy to see that these terms would preserve the
VEV structure of Eq. (\ref{vev}), they merely shift the vacuum expectation values given in Eq. (\ref{ui}) by small amounts.

\subsubsection{Realistic fermion masses}

Now we show how realistic fermion masses can be generated with $SO(3)_H$ family symmetry.  Since the Higgs doublets of MSSM
(denoted schematically as ${\bf 10}_H$ of $SO(10)$) are assumed to be $SO(3)_H$ singlets, additional fermions must exist with
GUT scale masses.  We assume as in Ref. \cite{babu} three pairs of ${\bf 16+ \overline{16}}$, denoted as ${\bf 16}_{A,B,C} +
{\bf \overline{16}}_{A,B,C}$.  These fields, which are all $SO(3)_H$ singlets, transform under the $Z_2 \times Z_2$ symmetry as ${\bf 16}_A + {\bf \overline{16}}_A:  (-+), {\bf 16}_B+ {\bf \overline{16}}_B: (+-)$ and ${\bf 16}_C + {\bf \overline{16}}_C: (--)$.  The most general renormalizable
Yukawa superpotential consistent with these symmetries is
\begin{eqnarray}
W_{\rm Yuk} &=& a_3 (\vec{\bf 16} \cdot \vec{A})\, {\bf \overline{16}}_A + a_2 (\vec{\bf 16} \cdot \vec{B})\, {\bf \overline{16}}_B +
a_1 (\vec{\bf 16} \cdot \vec{C})\, {\bf \overline{16}}_C  \nonumber \\
&+& \sum_{\alpha = A,B,C}M_\alpha\,{\bf \overline{16}}_\alpha {\bf 16}_\alpha + \sum_{\alpha = A,B,C} b_\alpha \, {\bf 16}_\alpha {\bf 16}_\alpha {\bf 10}_H~.
\label{Yuk2}
\end{eqnarray}

With the vacuum structure of Eq. (\ref{vev}), the couplings of Eq. (\ref{Yuk2}) would lead to a $3 \times 3$ mass matrix for each
type of fermion per family.  This matrix for the mixing of top quark can be written down as
\begin{eqnarray}
\left(\begin{matrix} t & t' & \overline{t}^{c \prime} \end{matrix}  \right) \left( \begin{matrix}0 & 0 & a_3 u_3 \\ 0 & b_3 v_u & M_A \\
    a_3 u_3 & M_A & 0\end{matrix}   \right) \left(\begin{matrix} t^c \\ t^{c \prime} \\ \overline{t}'       \end{matrix}   \right),
    \label{topmixing}
\end{eqnarray}
where ${\bf 16}_A \supset (t', \,t^{c \prime})$, and ${\bf \overline{16}}_A \supset (\overline{t}',\,\overline{t}^{c \prime})$.
The two heavy quarks can be integrated out,
which have a common mass given by $\sqrt{|a_3 u_3|^2 + |M_A|^2}$.  The top quark mass is then found to be
\begin{equation}
m_t = b_3 v_u \left(\frac{|a_3 u_3|^2}{|a_3 u_3|^2+|M_A|^2}  \right)~.
\label{top}
\end{equation}
In the limit of decoupled generations -- more about generation mixing later -- we can write down analogous expressions for the
lighter family fermion masses.  Thus we have
\begin{equation}
m_c = b_2 v_u \left(\frac{|a_2 u_2|^2}{|a_2 u_2|^2+|M_B|^2}  \right),~~~m_u = b_1 v_u \left(\frac{|a_1 u_1|^2}{|a_1 u_1|^2+|M_C|^2}  \right)~.
\end{equation}

For the top quark mass to be of order $0.5 \,v_u$ at the GUT scale (corresponding to $\tan\beta = 10$) \cite{tasi},
$|a_3 u_3| \sim |M_A|$ is required, since the Yukawa coupling
$|b_3|$ cannot exceed unity for perturbation theory to be valid.  This in turn would imply
that there is large mixing between ${\bf 16}_3$ and ${\bf 16}_A$.  Indeed, from Eq. (\ref{topmixing}) we can calculate this mixing.
The lighter top quark $\hat{t}$ and the heavy quark $\hat{t}'$ superfields are given by the combinations
\begin{equation}
\hat{t} = \frac{M_A^* t - (a_3 u_3)^* t'}{\sqrt{|a_3 u_3|^2 + |M_A|^2}},~~~~ \hat{t}' = \frac{M_A t' + (a_3 u_3) t }{\sqrt{|a_3 u_3|^2 + |M_A|^2}}~.
\end{equation}
All three families of scalars have a common soft SUSY breaking squared mass, denoted as $m_0^2$, by virtue of the $SO(3)_H$
symmetry.  The scalars from ${\bf 16}_A$ will have a different soft mass, denoted as $m_{16_A}^2$.  Due to the large mixing between
${\bf 16}_3$ and ${\bf 16}_A$, the soft mass of the light field $\hat{t}$ is given by
\begin{equation}
m^2_{\hat{t}} = m_0^2 + (m_{16_A}^2-m_0^2)\frac{|a_3 u_3|^2}{|a_3u_3|^2+|M_A|^2}~.
\label{split-t}
\end{equation}
Since $|a_3 u_3| \sim |M_A|$ from a fit to the top quark mass, we see that the soft mass for the SUSY partner of top is split by
order one from $m_0^2$.  Thus even with the {\bf 3} assignment of families under $SO(3)_H$, we obtain an effective low energy
description similar to a ${\bf 2+1}$ assignment.

The soft masses of the lighter generations will be given by relations analogous to Eq. (\ref{split-t}).  For example, the soft
mass of the strange squark is
\begin{equation}
m^2_{\hat{s}} = m_0^2 + (m_{16_B}^2-m_0^2)\frac{|a_2 u_2|^2}{|a_2u_2|^2+|M_B|^2}~.
\label{split-s}
\end{equation}
In this case however, $|a_2 u_2| \ll |M_B|$ can be chosen.  For $b_2 = 1$, a fit to the strange quark mass $m_s(M_{\rm GUT}) = 13$ MeV
\cite{tasi} (corresponding to $\tan\beta = 10$) would set $|a_2 u_2|^2/|M_B|^2 \simeq 7.5 \times 10^{-4}$.
The strength of FCNC arising from
the splitting of $\tilde{s}$ and $\tilde{d}$ is   $|(\delta^d_{LL})_{12} (\delta^d_{RR})_{12}|^{1/2}$ $\sim 1.5 \times 10^{-4}$, obtained from $|a_2 u_2|^2/|M_B|^2 \,\theta_C$.  This is
within the bounds quoted in  Eq. (\ref{limit}). Note that the mixing of $\tilde{d}$ with ${\bf 16}_C$ is negligibly small and has been
ignored in this estimate.

With the $Z_2 \times Z_2$ symmetry and the structure of the VEVs of Eq. (\ref{vev}) the model would not have any generation mixing.
This can be cured by assigning $Z_2 \times Z_2$ charge of $(--)$ to certain GUT multiplets, such as the ${\bf 45}_H$.  This would
allow inter-generational couplings of the type ${\bf 16}_A {\bf \overline{16}}_B {\bf 45}_H$ in the heavy sector \cite{babu}.
If a true GUT symmetry is employed,
such couplings would also help split the quark masses from the lepton masses which would otherwise yield unacceptable mass relations
such as $m_s = m_\mu$.  Assigning such a $Z_2 \times Z_2$ charge to GUT multiplets will not upset the VEV structure of Eq. (\ref{vev}),
and the $D$-terms of $SO(3)_H$ will remain zero.  The model can explain SUSY CP problem via spontaneous CP violation as in the case
of $SU(2)_H$ model.  A singlet field $X$ with complex VEV can couple to the vectors $(\vec{A},\,\vec{B},\,\vec{C})$ through terms
such as $\vec{A} \cdot \vec{A} \,X$ etc.

\subsection{sMSSM from permutation symmetry \boldmath{$S_3$}}

In this section and the next we suggest and analyze flavor symmetries based on non-Abelian discrete groups that solve the
SUSY flavor violation problem.  Being discrete, these models do not suffer from any $D$-term issues, as there are no
$D$-terms associated with discrete symmetries.  Realistic fermion masses must be generated, which requires
analyzing the breaking of the discrete symmetry in some detail.  These symmetries will be shown to lead to sMSSM phenomenology.

The simplest non-Abelian symmetry is $S_3$, the permutation symmetry acting on three letters.  ($S_3$ is isomorphic with the dihedral group $D_3$.)  It is an order 6 group with irreducible representations ${\bf 1}, {\bf 1'}$ and
${\bf 2}$.  The ${\bf 1}$ and ${\bf 1'}$ belong to an $S_2$ subgroup (isomorphic with $Z_2$).   The Kronecker products of the irreps are given as
\begin{equation}
{\bf 1'} \times {\bf 1'} = 1,~~~{\bf 2} \times {\bf 2} = {\bf 1} + {\bf 1'} + {\bf 2}~.
\end{equation}
In a certain basis, the Clebsch-Gordon coefficients for the multiplication ${\bf 2} \times {\bf 2}$ is given as
\begin{eqnarray}
\left( \begin{matrix} a \\ b \end{matrix} \right) \times \left( \begin{matrix} a' \\ b' \end{matrix} \right) = a a' + b b'
\sim {\bf 1},~~
a b' - b a' \sim {\bf 1'},~~
\left(\begin{matrix} a b' + b a' \\ a a' - b b'  \end{matrix}  \right) \sim {\bf 2}~.
\end{eqnarray}

Under $S_3$, the three families of ${\bf 16}$ are assigned as ${\bf 2} + {\bf 1}$, with the ${\bf 1}$ identified as the third family.
This assignment is similar to the case of $SU(2)_H$ model discussed earlier.  Here there will be no restriction from the vanishing
of $D$-terms, unlike in the $SU(2)_H$ model.
The Higgs field ${\bf 10}_H$ is a singlet of $S_3$.  (It should be  emphasized that we are not constructing explicit $SO(10)$ GUT models, but our construction will be compatible with $SO(10)$.)  In the $S_3$ symmetric limit, only  the third family would acquire a mass, through the
superpotential coupling $W \supset Y_3 {\bf 16}_3 {\bf 16}_3 {\bf 10}_H$.  The lighter generation masses would arise from higher
dimensional operators after $S_3$ symmetry breaks spontaneously.

It will turn out that $S_3$ symmetry has to be appended with an Abelian symmetry which we choose to be $Z_3$.  Thus the full symmetry
of the model is $S_3 \times Z_3$.  There are two purposes for
the $Z_3$.  First, it would prevent an $S_3$-invariant Yukawa coupling $({\bf 16}_1 {\bf 16}_1 + {\bf 16}_2 {\bf 16}_2)\,{\bf 10}_H$,
which would induce a common mass for the first two families at the same order as the third family mass.  The $Z_3$ charge assignment
of $({\bf 16}_1,\,{\bf 16}_2): \omega^2$, ${\bf 16}_3: 1$, and ${\bf 10}_H: 1$ where $\omega^3 = 1$, would prevent such a term.
Second, $S_3$ alone would allow a K\"ahler potential term leading to the Lagrangian
\begin{equation}
{\cal L} \supset \int d^4\theta\, c_{ijk} (q_i^\dagger q_j S_k) \frac{|Z|^2}{M_{\rm Pl}^3}
\end{equation}
where $S$ is a SM singlet field which is a doublet of $S_3$ that is needed for its spontaneous breaking.
Here $c_{121}=c_{211}=c_{112}=-c_{222} \neq 0$, while all other $c_{ijk}$ vanish.  This Lagrangian term would induce an off-digaonal squark
mixing of order $m_{\rm SUSY}^2 \left\langle S \right\rangle/M_{\rm Pl} \sim 10^{-2}\,m_{\rm SUSY}^2$, which would be in violation
of the limit of Eq. (\ref{limit}).  With the $Z_3$ symmetry, this term will not be allowed in the Lagrangian, as the filed $S$ would
carry $Z_3$ charge of $\omega$.  The leading correction to the K\"ahler potential with the $Z_3$ symmetry is
\begin{equation}
{\cal L} \supset \int d^4\theta\, c_{ijkl} (q_i^\dagger q_j S_k^\dagger S_l) \frac{|Z|^2}{M_{\rm Pl}^4}
\end{equation}
which would induce off-diagonal squark mixing of order  $m_{\rm SUSY}^2 \left\langle S \right\rangle^2/M^2_{\rm Pl} \sim 10^{-4}\,m_{\rm SUSY}^2$,
which is acceptable.

The full $S_3 \times Z_3$ assignment of matter fields of the model is as follows:
\begin{eqnarray}
&~& ({\bf 16}_1,\, {\bf 16}_2): (2, \omega^2);~~~~{\bf 16}_3: (1,1);~~~~{\bf 10}_H: (1,1);\nonumber \\
&~& S = \left(\begin{matrix} S_1 \\ S_2 \end{matrix}  \right):
(2, \omega);~~~~\eta: (1, \omega);~~~~ \overline{\eta}: (1,\omega^2)~.
\end{eqnarray}
Here $S,\,\eta,\,\overline{\eta}$ are SM and $SO(10)$ singlet fields with $\eta+\overline{\eta}$ needed for $Z_3$ symmetry breaking.

The superpotential relevant for high scale symmetry breaking involving the fields $S,\,\eta,\,\overline{\eta}$ is given by
\begin{equation}
W = \mu_\eta\, \eta \overline{\eta} + \lambda_1 \,\eta^3 + \lambda_2\, \overline{\eta}^3 + \lambda_3 \,\eta\,(S_1^2+S_2^2) + \lambda_4 \,S_2(3S_1^2-S_2^2)~.
\end{equation}
This superpotential leads to the following symmetry breaking minima in the SUSY limit:
\begin{eqnarray}
\left\langle S \right \rangle = u\,\left(\begin{matrix} 0 \\ 1 \end{matrix} \right) ~~~~{\rm or}~~~~ u \left(\begin{matrix} \pm \sqrt{3}
\\ 1 \end{matrix}  \right)
\end{eqnarray}
with
\begin{equation}
u = \frac{2 \lambda_3 \lambda_4^{1/3} \omega^2 \mu_\eta}{\lambda_2^{1/3} \left(4 \lambda_3^3 + 27 \lambda_1 \lambda_4^2\right)^{2/3}}~~~~
{\rm or} ~~~~ u = \frac{\pm \lambda_3 \lambda_4^{1/3} \mu_\eta}{\lambda_2^{1/3} \left(4 \lambda_3^3 + 27 \lambda_1 \lambda_4^2\right)^{2/3}}~.
\end{equation}
In each solution a separate $S_2$ subgroup of $S_3$ is preserved, even with $\eta$ and $\overline{\eta}$ acquiring VEVs.  With such an
unbroken subgroup, one of the families will decouple from the rest.

There are two simple ways of breaking the $S_3$ symmetry completely.  A straightforward way would involve introducing a second
$S'(2,\omega)$ field.  If the cross couplings between $S$ and $S'$ are ignored, a solution of the type
\begin{eqnarray}
\left\langle S \right \rangle = u\,\left(\begin{matrix} 0 \\ 1 \end{matrix} \right); ~~~~~~\left\langle S' \right \rangle = u' \left(\begin{matrix}  \sqrt{3}
\\ 1 \end{matrix}  \right)
\end{eqnarray}
can be chosen.  Since $\left\langle S \right \rangle$ and $\left\langle S' \right \rangle$ break $S_3$ to separate $S_2$ subgroups, their
combined effect would be to break $S_3$ completely.  Once mixed couplings of the type $S^2S'$ and $S S'^2$ are turned on, the VEV structure
of $\left\langle S \right \rangle$ and $\left\langle S' \right \rangle$ would be parameter dependent, suggesting complete breakdown of $S_3$.

A second way to break $S_3$ completely without introducing the $S'$ field is to assign a GUT multiplet such as ${\bf 45}_H$ as a ${\bf 1'}$ of $S_3$,
which is not very restrictive.  Once ${\bf 45}_H$ acquires a VEV, the unbroken $S_2$ subgroup will break completely.  In both cases we
shall see that realistic fermion masses can be generated.

\subsubsection{Realistic fermion masses}

The Yukawa superpotential of the $S_3 \times Z_3$ model can be written down as
\begin{eqnarray}
W_{\rm Yuk} &=& a_3 {\bf 16}_3 {\bf 16}_3 {\bf 10}_H + \frac{a_2}{M_*} {\bf 16}_i S_i {\bf 16}_3 {\bf 10}_H + \frac{a_1}{M_*^2}({\bf 16}_1 {\bf 16}_2 - {\bf 16}_2
{\bf 16}_1)\, {\bf 45}_H \,{\bf 10}_H \,\overline{\eta} \nonumber \\
 &+& \frac{a_1'}{M_*^2} \left(4 \times{\bf 16}_1 {\bf 16}_2\,S_1\, S_2 + ({\bf 16}_1 {\bf 16}_1 - {\bf 16}_2 {\bf 16}_2)(S_1^2-S_2^2)\right)\,{\bf 10}_H+ ...
 \label{Yuk3}
\end{eqnarray}
If $S'$ field is not utilized, ${\bf 45}_H$ should be a ${\bf 1'}$ under $S_3$.  In this case, the fermion mass matrix would have the same
form as Eq. (\ref{Mf1}) obtained with $SU(2)_H$ symmetry, along with a small equal and opposite diagonal entries in the (1,1) and (2,2) locations
proportional to the coupling $a_1'$.  This statement holds if $\left\langle S \right \rangle \propto (0,\,1)^T$ is used.  If a second $S_3$ doublet
$S'$ is present, in the couplings of Eq. (\ref{Yuk3}) wherever $S$ appears it may be replaced by $S'$ as well.  In either case we find that
realistic fermion masses can be generated with a qualitative understanding of the mass hierarchy. Here again  superpotential in Eq. (\ref{Yuk3}) can  give a desirable framework  for $t-b-\tau$ Yukawa coupling unification  \cite{yukawaUn, Ajaib:2013zha}.

As for the SUSY breaking mass parameters, the $S_3$ doublet $({\bf 16}_1,~{\bf 16}_2)$ will have a common mass $m_0$,
while ${\bf 16}_3$ will have a separate soft mass $m_{3(0)}$.  There is no reason, based on symmetries, to assume any hierarchy between
$m_0$ and $m_{3(0)}$, so in our phenomenological analysis we take them to be of the same order.  Also, there is
no symmetry reason that would set $m_0 = m_{3(0)}$, which is usually assumed in cMSSM. As in the case of $SU(2)_H$ model, SUSY
induced flavor violation arising from $m_0 \neq m_{3(0)}$ is under control in this $S_3 \times Z_3$ setup.

\subsection{ sMSSM from \boldmath{$A_4$} flavor symmetry}

In this section we develop a model based on $A_4$ flavor symmetry that solves the SUSY flavor violation problem
and also sheds some light on the fermion mass puzzle.  $A_4$ is the simplest group that contains a triplet representation.
It is the symmetry group of a regular tetrahedron.  It can be also viewed as the group of even permutation of four letters.
The irreducible representations of this order 12 group fall into ${\bf 1,\,1',\,1'',\,3}$, with the $({\bf 1,\,1',\,1''})$ forming
a $Z_3$ subgroup.  The nontrivial Kronecker products are given by
\begin{eqnarray}
{\bf 1'} \times {\bf 1''} = {\bf 1},~~{\bf 1'} \times {\bf 1'} = {\bf 1''},~~{\bf 1''} \times {\bf 1''} = {\bf 1'},~~{\bf 3} \times {\bf 3}
= {\bf 1}_s + {\bf 1'}_s + {\bf 1''}_s + {\bf 3}_s + {\bf 3}_a~.
\end{eqnarray}
In a certain basis, the product of two triplets has the decomposition $(a_1,\,a_2,\,a_3) \times (b_1,\,b_2,\,b_3) =
(a_1 b_1 + a_2 b_2 + a_3 b_3) \sim {\bf 1}$; $(a_1 b_1 + \omega^2 a_2 b_2 + \omega a_3 b_3) \sim {\bf 1'}$; $(a_1 b_1 + \omega a_2 b_2 + \omega^2
a_3 b_3) \sim {\bf 1''}$; $(a_2 b_3 + a_3 b_2,\,a_3b_1 + a_1 b_3,\,a_1 b_2 + a_2 b_1)\sim {\bf 3}_s$ and
$(a_2 b_3 - a_3 b_2,\,a_3b_1 - a_1 b_3,\,a_1 b_2 - a_2 b_1)\sim {\bf 3}_a$, where $\omega = e^{2 i \pi/3}$.

The $A_4$ model we propose is very similar to the $SO(3)_H$ model discussed earlier.  The main difference is that there is no
constraint arising from the $D$-terms in the present case.  SUSY flavor violation arising from higher dimensional operators
should be sufficiently suppressed.  This is achieved by supplementing $A_4$ by a $Z_2 \times Z_2$ symmetry, as in the case
of $SO(3)_H$ model.  The three families are assigned to {\bf 3} of $A_4$ and they transform trivially under $Z_2 \times Z_2$.
That is, they belong to $({\bf 16}, {\bf 3})(++)$ under $(SO(10) \times A_4) \times (Z_2 \times Z_2)$.

$A_4$ symmetry breaking is achieved by three $A_4$ triplets which are singlets of the SM and $SO(10)$.  We denote them
as $({\vec A},\,\vec{B},\,\vec{C})$.  These fields transform as $\vec{A}: (-+),\,\vec{B}: (+-),\,\vec{C}: (--)$ under
$Z_2 \times Z_2$. The superpotential of these fields can be written down as
\begin{equation}
W = \frac{\mu_A}{2} \vec{A} \cdot \vec{A} + \frac{\mu_B}{2} \vec{B} \cdot \vec{B} + \frac{\mu_C}{2} \vec{C} \cdot \vec{C} +
\lambda_1 \,\vec{A} \times \vec{B} \cdot \vec{C} + \lambda_2 \, (a_1 b_2 c_3 + a_2 b_3 c_1 + a_3 b_1 c_2)
\end{equation}
where $\vec{A}_i = a_i$ etc have been used.  This potential will reduce to the one of Eq. (\ref{W2}) corresponding to $SO(3)_H$
symmetry in the absence of the $\lambda_2$ term which ensures the absence of Goldstone bosons.  The soft SUSY breaking Lagrangian
is given by
\begin{eqnarray}
V_{\rm soft} &=& m_A^2 \vec{A}^\dagger\cdot  \vec{A} +  m_B^2 \vec{B}^\dagger \cdot \vec{B} +  m_C^2 \vec{C}^\dagger \cdot \vec{C} +\left \{
\frac{B_A \mu_A}{2} \vec{A} \cdot \vec{A} + \frac{B_B \mu_B}{2} \vec{B} \cdot \vec{B}
+ \frac{B_C \mu_C}{2} \vec{C} \cdot \vec{C} \right. \nonumber \\
&~&\left.  + A_{\lambda_1} \lambda_1 \,\vec{A} \times \vec{B} \cdot \vec{C} + A_{\lambda_2}\lambda_2 \, (a_1 b_2 c_3 + a_2 b_3 c_1 + a_3 b_1 c_2) + h.c.\right\}
\end{eqnarray}
Note that there is no $D$-term contributions in this model, so that the full potential is given as $V = V_F + V_{\rm soft}$. This potential
admits a vacuum structure of the form
 \begin{eqnarray}
 \left\langle \vec{A} \right \rangle = \left(\begin{matrix} 0 \\ 0 \\ u_3 \end{matrix}  \right),~~~~
 \left\langle \vec{B} \right \rangle = \left(\begin{matrix} 0 \\ u_2 \\ 0 \end{matrix}  \right),~~~~
 \left\langle \vec{C} \right \rangle = \left(\begin{matrix} u_1 \\ 0 \\ 0 \end{matrix}  \right)~.
 \label{vev2}
 \end{eqnarray}
 This is identical to the VEV structure in the $SO(3)_H$ model, even though there are new couplings in the $A_4$ model.

This vacuum structure is stable against higher dimensional operators.  For example, there are four couplings of the
form $(A)^2 (B)^2$ in the superpotential, suppressed by one inverse power of the Planck mass.  They are
$(a_1^2+a_2^2+a_3^2)(b_1^2+b_2^2+b_3^2)$; $(a_1^2+ \omega^2 a_2^2+ \omega a_3^2)(b_1^2+ \omega b_2^2+ \omega^2 b_3^2)$;
$(a_1^2+ \omega a_2^2+ \omega^2 a_3^2)(b_1^2+ \omega^2 b_2^2+ \omega b_3^2)$; and $(a_1 a_2 b_1 b_2+a_1 a_3 b_1 b_3 + a_2 a_3 b_2 b_3)$.
None of these terms would upset the VEVs of Eq. (\ref{vev2}).

The model cures the SUSY flavor violation problem rather well.  For example, the leading K\"ahler potential
correction to SUSY FCNC arises from
 \begin{equation}
{\cal L} \supset \int d^4\theta  (\vec{Q}^\dagger \cdot \vec{B}) (\vec{B}^\dagger \cdot \vec{Q}) \frac{|Z|^2}{M_{\rm Pl}^4}~.
\label{kahler4}
\end{equation}
This would lead to  off-diagonal $\tilde{d}-\tilde{s}$ mixing of order $m_{\rm SUSY}^2 |u_2|^2/M_{\rm Pl}^2 \sim 10^{-4}\,m_{\rm SUSY}^2$, well within
limits shown in Eq. (\ref{limit}).

Fermion mass generation in the $A_4$ model parallels that for the $SO(3)_H$ model, so we can be brief here.  Three pairs of
${\bf 16} + {\bf \overline{16}}$ fermion superfields are introduced which are $A_4$ singlets, but carry $Z_2 \times Z_2$
charges of $(-+),\,(+-),\,(--)$.  The mixing of $({\bf 16,3})$ with these fermions induce realistic fermion masses.  The mixing
of ${\bf 16}_3$ with ${\bf 16}_A$ is of order unity, so that the soft mass of ${\bf 16}_3$ will be split by order one from those
of $({\bf 16}_1,\,{\bf 16}_2)$ which remain nearly degenerate.  The resulting SUSY phenomenology is that of sMSSM.

\section{Phenomenological analysis of sMSSM}

In this section we perform a detailed phenomenological analysis of sMSSM motivated in the previous sections.
As noted in the introduction, the parameter set
of sMSSM contains three more variables compared to cMSSM.  We perform a detailed scan of this enlarged parameter space
and look for consistent solutions which satisfy the following phenomenological requirements:  (i) Successful radiative electroweak
symmetry breaking, (ii) the lightest supersymmetric particle must be electrically neutral, (iii) all constraints from
$B$ meson physics must be satisfied, (iv) experimental limits on SUSY particle must be satisfied, and (v) the lightest Higgs boson
must have a mass in the range of 124-126 GeV.  In addition, we search for regions in the parameter space which would provide
the correct amount of relic abundance of LSP dark matter.  We also present our results where the relic dark matter abundance
obeys $\Omega h^2 < 1$.  These points are likely to lead to the needed dark matter abundance, since for typical points
$\Omega h^2 \gg 1$, and it is reduced to lower values primarily due to co-annihilation effects and/or resonance effects
that enhance the annihilation rate.  A finer scan around the points satisfying $\Omega h^2 < 1$ is likely to give the needed
abundance.

At the Lagrangian level, the sMSSM parameter set is $\{m_{{0}_{(1,2)}},\, m_{{0}_{(3)}},\, M_{1/2},\,m_{H_u}^2,\,m_{H_d}^2,\,
\mu$, $A_0,\,B\}$.  Of these eight parameters, $m_{H_u}^2,\,m_{H_d}^2$ and $B$
can be traded for $M_Z$, $m_A$ and $\tan\beta$, which would lead to the seven parameter set shown in Eq. (\ref{para}).  The procedure
we adopt is described in detail, followed by a discussion of our numerical results.

\subsection{Scanning Procedure, Parameter Space and Experimental Constraints\label{constraintsSection}}

We employ the ISAJET~{7.84} package~\cite{ISAJET}  to perform random
scans over the fundamental parameter space. In this package, the weak scale values of gauge couplings and the
third generation Yukawa
couplings are evolved to $M_{\rm GUT}$ via the MSSM RGEs in the $\overline{DR}$ regularization scheme.
We define $M_{\rm GUT}$ as the meeting point of $\alpha_1$ and $\alpha_2$, which is obtained iteratively.
The value of $M_{\rm GUT}$ is found to vary between $1.4 \times 10^{16}$ GeV and $3 \times 10^{16}$ GeV.
We do not strictly enforce the unification condition $\alpha_3=\alpha_2$ at $M_{\rm
GUT}$, since a few percent deviation from unification can be
assigned to unknown GUT-scale threshold
corrections~\cite{Hisano:1992jj}.
The deviation between $g_1=g_2$ and $g_3$ at $M_{\rm GUT}$ is no
worse than $3-4\%$.
For simplicity,  we do not include the Dirac neutrino Yukawa coupling
in the RGEs, whose contribution is expected to be small.

The various boundary conditions are imposed at
$M_{\rm GUT}$ and all the SSB
parameters, along with the gauge and Yukawa couplings, are evolved
back to the weak scale $M_{\rm Z}$.
In the evolution of Yukawa couplings the SUSY threshold
corrections~\cite{Pierce:1996zz} are taken into account at the
common scale $M_{\rm SUSY}= \sqrt{m_{{\tilde t}_L}m_{{\tilde t}_R}}$,
where $m_{{\tilde t}_L}$ and $m_{{\tilde t}_R}$
denote the masses of the third generation left and right-handed stop quarks.
The entire parameter set is iteratively run between $M_{\rm Z}$ and $M_{\rm
GUT}$ using the full 2-loop RGEs until a stable solution is
obtained. To better account for leading-log corrections, one-loop
step-beta functions are adopted for the gauge and Yukawa couplings, and
the SSB parameters $m_i$ are extracted from RGEs at multiple scales
$m_i=m_i(m_i)$. The RGE-improved 1-loop effective potential is
minimized at $M_{\rm SUSY}$, which effectively
accounts for the leading 2-loop corrections. Full 1-loop radiative
corrections are incorporated for all sparticle masses.

An approximate error of around 2 GeV is expected in the estimate of the Higgs boson mass in Isajet
which largely arises from theoretical uncertainties in {the calculation} of the minimum of the
scalar potential, and to a lesser extent from experimental uncertainties in the values
for $m_t$ and $\alpha_s$.

An important constraint on the parameter space arises from limits on the cosmological abundance of stable charged
particles  \cite{Beringer:1900zz}. This excludes regions in the parameter space
where  charged SUSY particles  become
the lightest supersymmetric particle (LSP). We accept only those
solutions for which one of the neutralinos is the LSP and saturates
the WMAP  bound on relic dark matter abundance.

We have performed random scans for the following parameter range:
\begin{align}
0 \leq m_{{0}_{(1,2)}}  \leq 3\,\rm {TeV} \nonumber \\
0 \leq m_{{0}_{(3)}}  \leq 3\,\rm {TeV} \nonumber \\
0\leq   M_{1/2} \leq 3\, \rm{TeV} \nonumber \\
-5\leq  {A_0}  \leq 5 \,\rm {TeV} \nonumber \\
2\leq \tan\beta \leq 20 \nonumber \\
0\leq \mu \leq 3 \,\rm {TeV}  \nonumber \\
0\leq m_{A} \leq 3 \,\rm {TeV}  \nonumber \\
\mu>0, m_t=173.3 GeV
\label{parameterRange}
\end{align}
{where $m_{0_{(1,2)}}$ is the SSB mass parameter for the first two generations, while $m_{0_{(3)}}$ is for the third generation of sfermions. $M_{1/2}$ is the SSB gaugino mass, $A_{0}$ is the SSB trilinear scalar interaction coupling. $\mu$ and $m_{A}$ are  bilinear Higgs mixing term and mass of the CP-odd Higgs boson respectively. In contrast to other parameters, $\mu$ and $m_{A}$ values are set at low scale. Besides, we} set $m_t = 173.3\, {\rm GeV}$  \cite{:1900yx}. Note that $m_b(m_Z)=2.83$ GeV, which is hard-coded into ISAJET.  Note that all SUSY breaking fundamental parameters have
magnitudes less than about 3 TeV in our scan (except for $A_0$ which is somewhat larger),
which would make most of the sparticles to be within reach of the LHC.  Such a range would also
imply that fine tuning in the Higgs boson mass is not very severe.  We have confined our analysis to small and moderate values of
$\tan\beta$ for simplicity.

In scanning the parameter space, we employ the Metropolis-Hastings
algorithm as described in \cite{Belanger:2009ti}. The data points
collected all satisfy
the requirement of radiative electroweak symmetry breaking (REWSB),
with the neutralino {being the LSP in each case}. After collecting the data, we impose
the mass bounds on all the particles \cite{ATLAS,CMS,Nakamura:2010zzi} and use the
IsaTools package~\cite{Baer:2002fv}
to implement the various phenomenological constraints. We successively apply the  experimental constraints presented in Table 1 on the data that
we acquire from ISAJET:

\begin{table}[h]\centering
\begin{tabular}{rlc}
$   124\, {\rm GeV} \leq m_h \leq126$ \,{\rm GeV}~~~&\cite{:2012gk,:2012gu}&
\\
$0.8\times 10^{-9} \leq{\rm BR}(B_s \rightarrow \mu^+ \mu^-)
  \leq 6.2 \times10^{-9} \;(2\sigma)$~~~&\cite{BsMuMu}&
\\
$2.99 \times 10^{-4} \leq
  {\rm BR}(b \rightarrow s \gamma)
  \leq 3.87 \times 10^{-4} \; (2\sigma)$~~~&\cite{Amhis:2012bh}&
\\
$0.15 \leq \frac{
 {\rm BR}(B_u\rightarrow\tau \nu_{\tau})_{\rm MSSM}}
 {{\rm BR}(B_u\rightarrow \tau \nu_{\tau})_{\rm SM}}
        \leq 2.41 \; (3\sigma)$~~~&\cite{Asner:2010qj}&
\\
$0.1088 \le \Omega_d h^2 \leq 0.1217$~~~&\cite{WMAP9}&
\end{tabular}
\caption{Various phenomenological constraints implemented { in} our study.}
\label{table1}
\end{table}
\begin{figure}
\centering
\subfiguretopcaptrue
\subfigure{
\includegraphics[width=6cm]{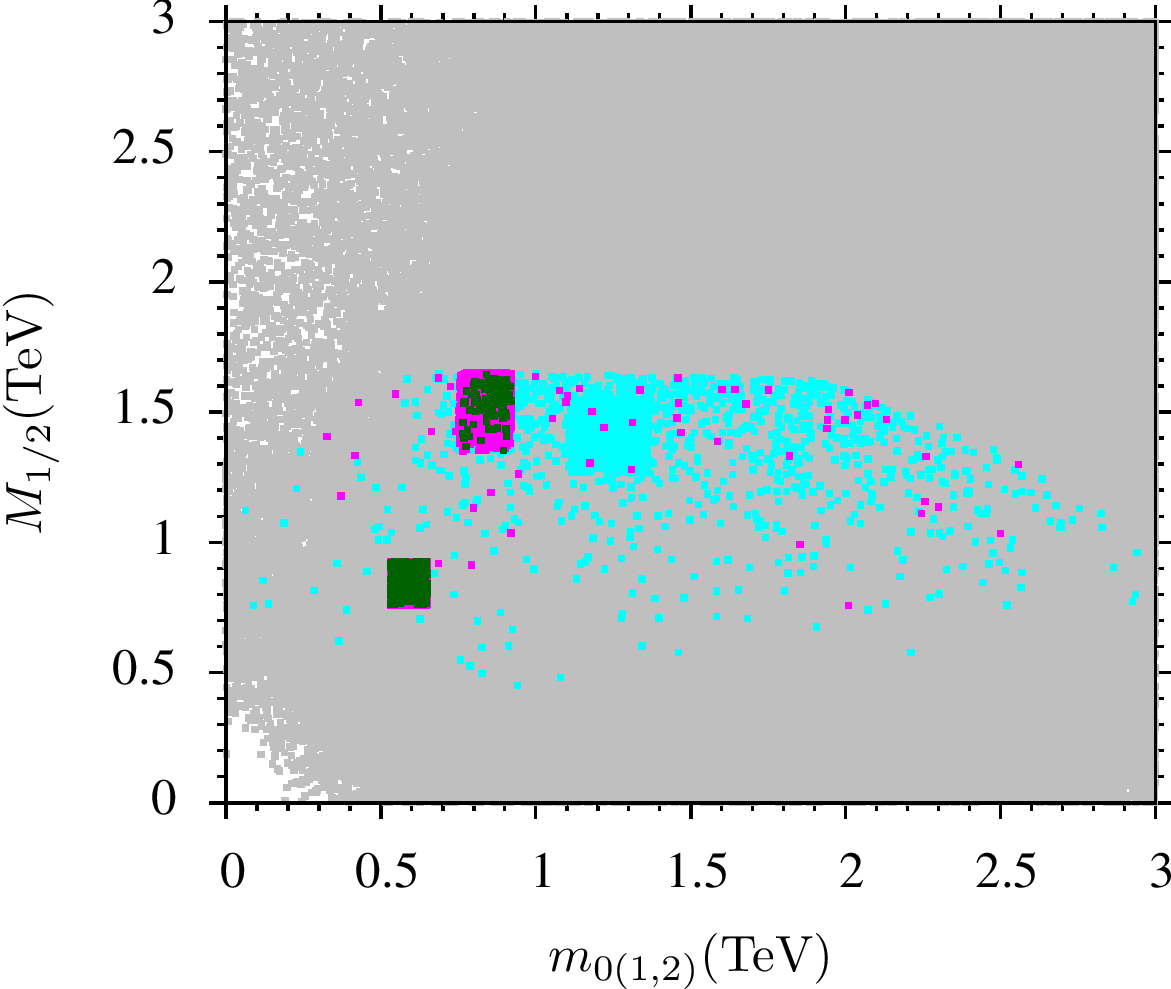}
}
\subfigure{
\includegraphics[width=6cm]{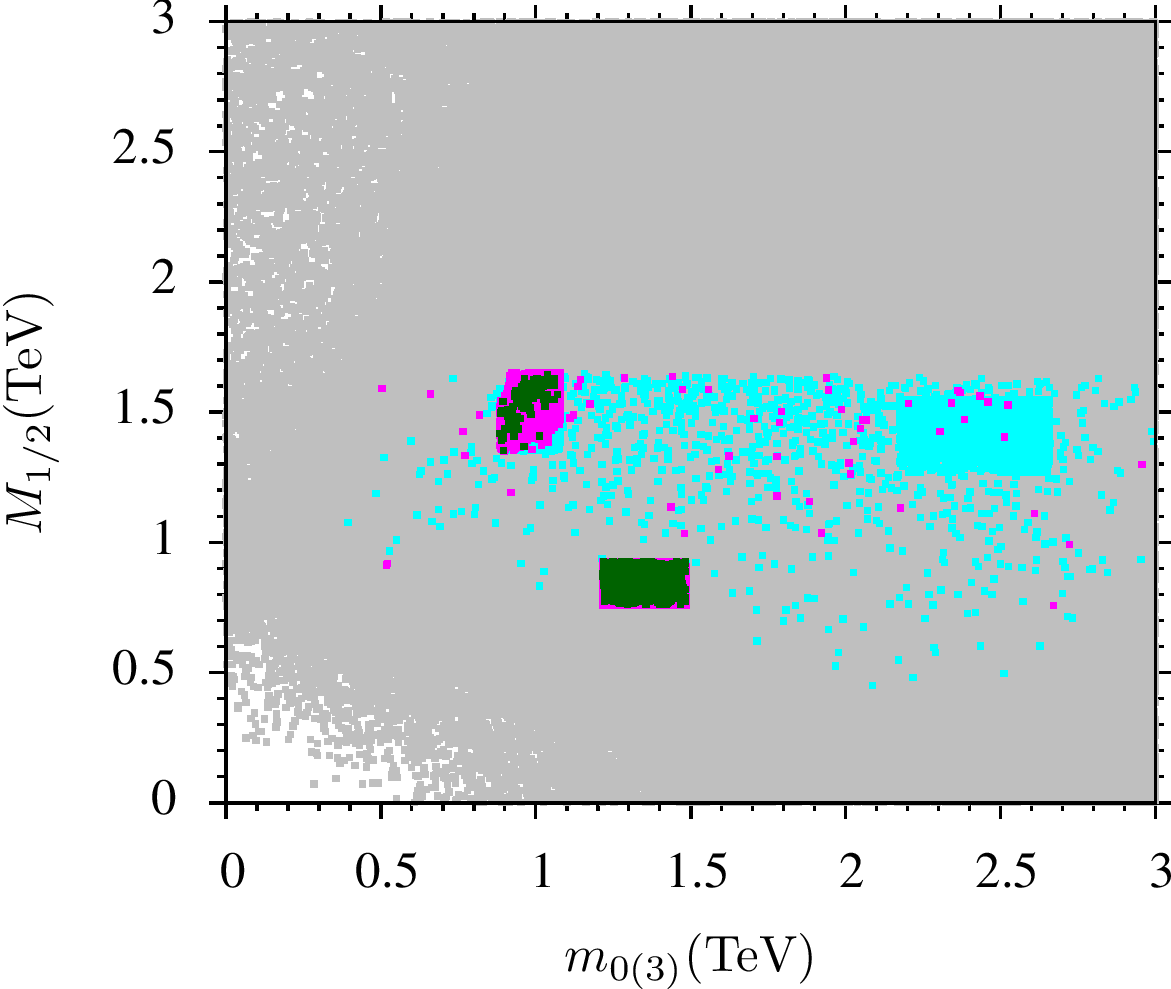} }
\subfigure{
\includegraphics[width=6cm]{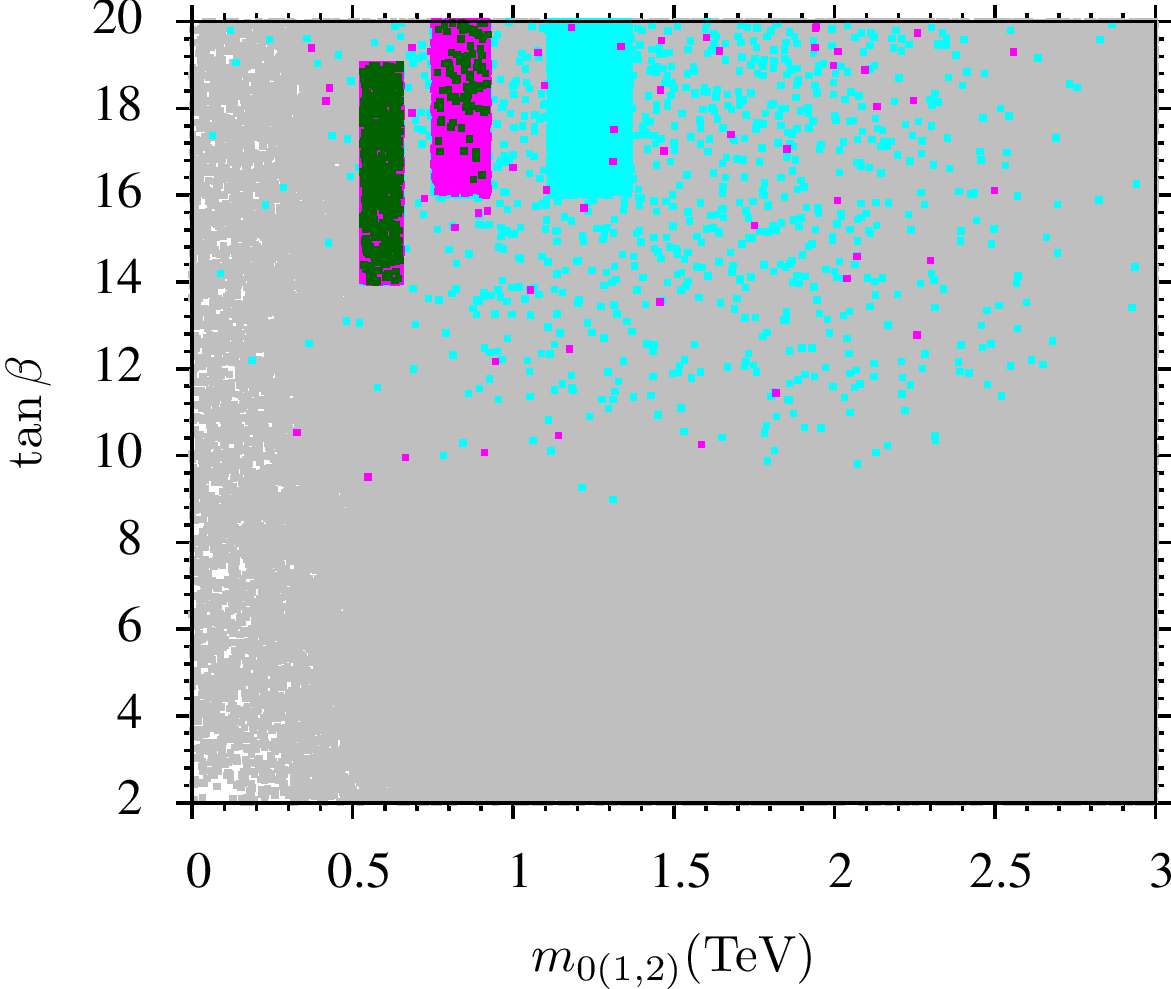}
}
\subfigure{
\includegraphics[width=6cm]{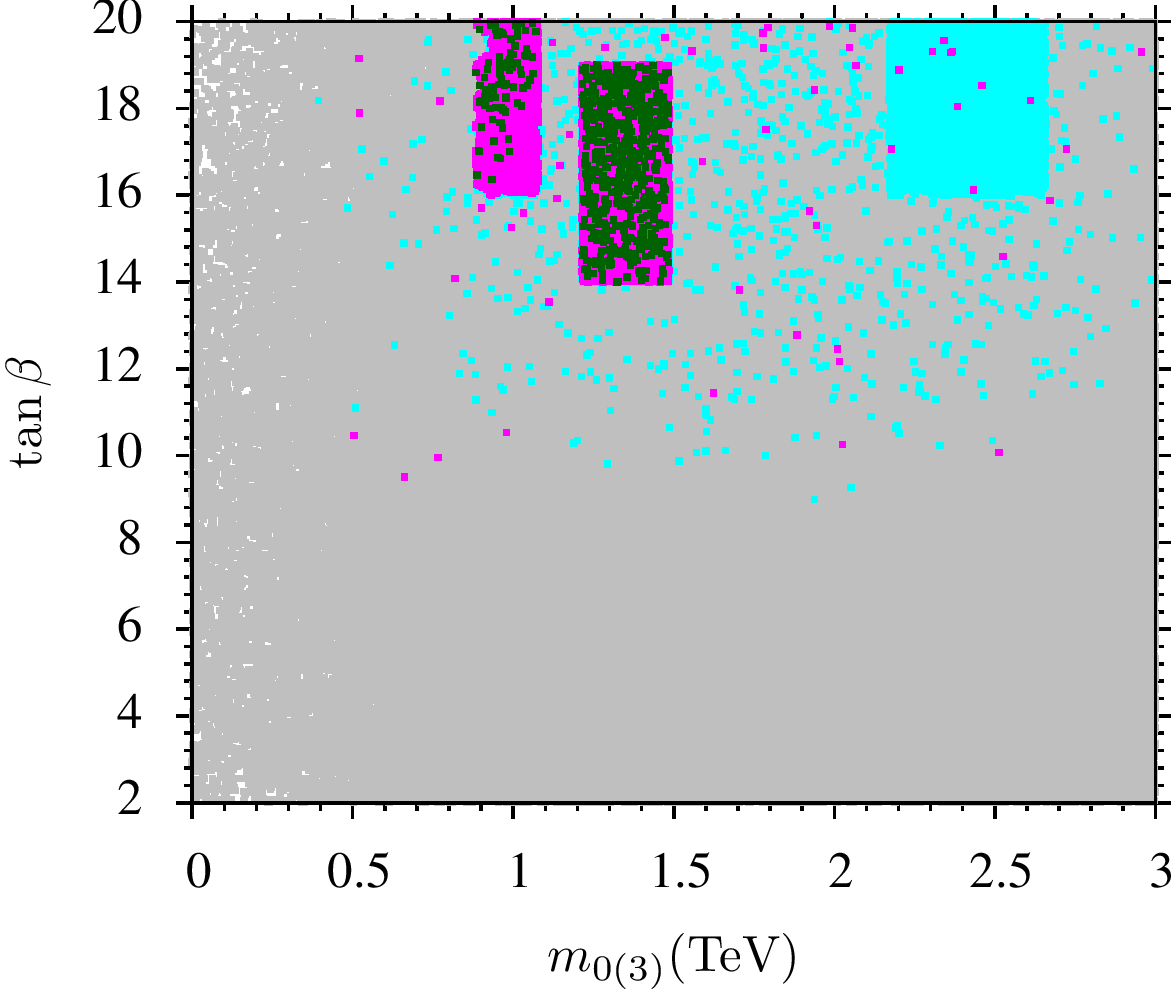} }
\subfigure{
\includegraphics[width=6cm]{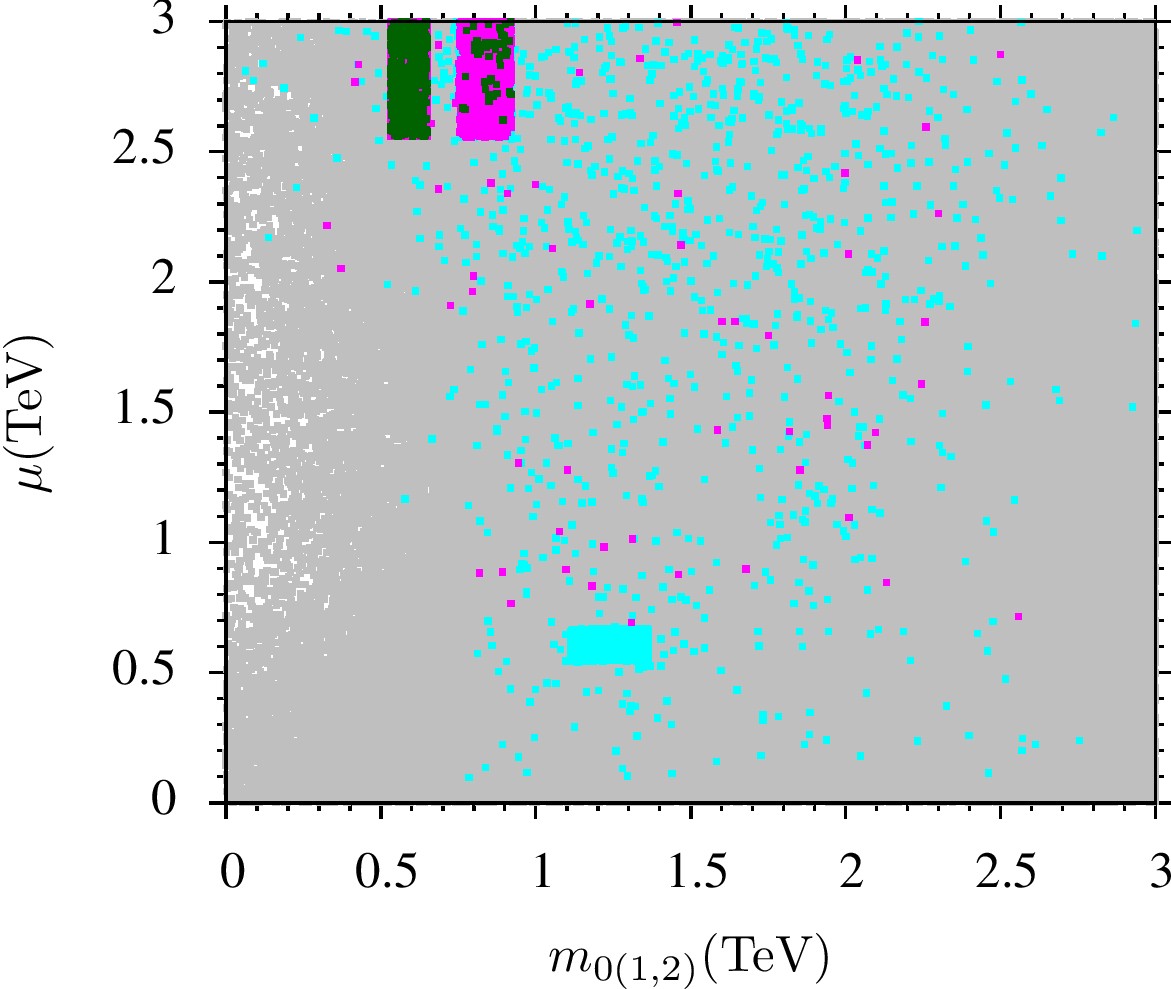}}
\subfigure{
\includegraphics[width=6cm]{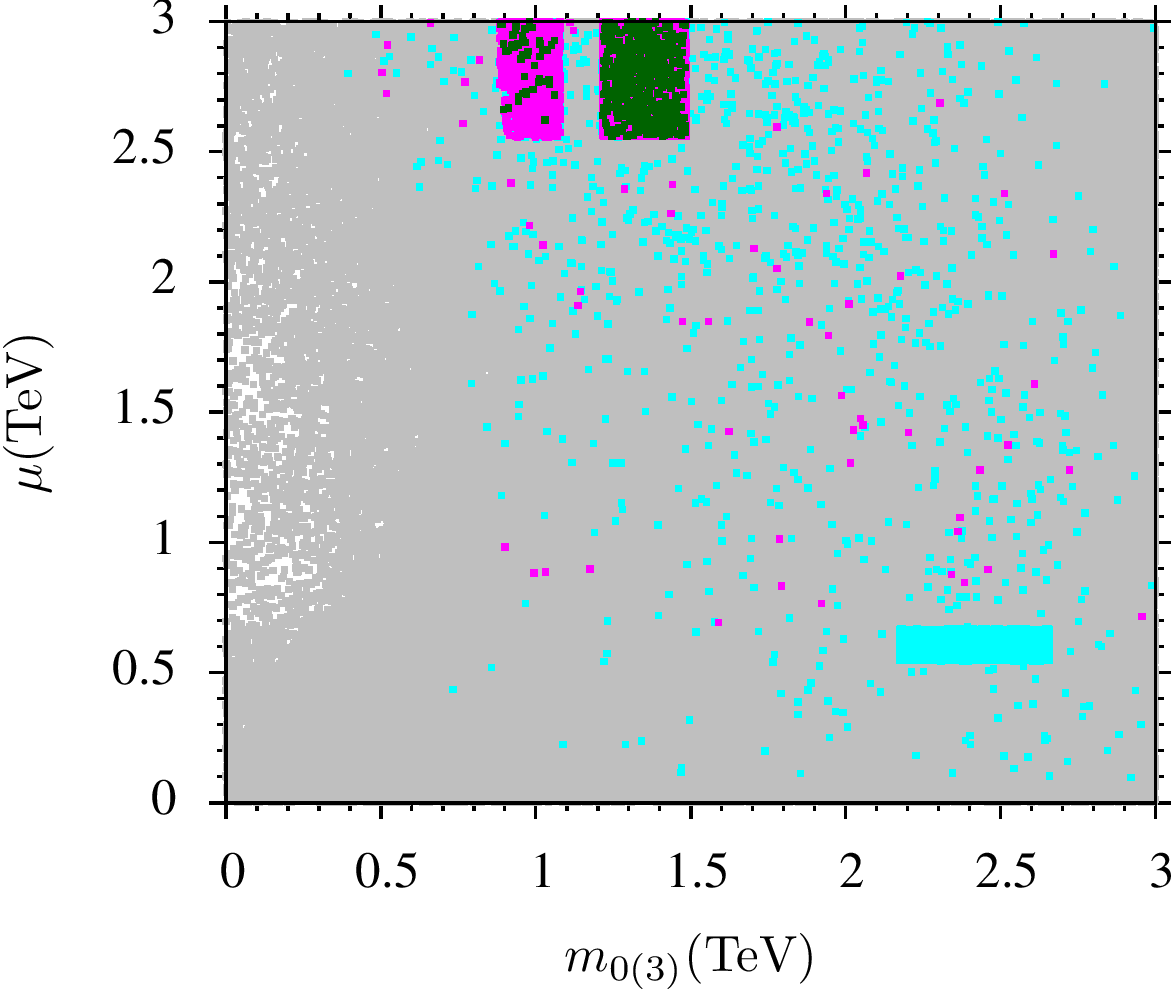}}
\caption{Plots in $m_{0_{(1,2)}}-M_{1/2}$, $m_{0_{(3)}}-M_{1/2}$, $m_{0_{(1,2)}}-\tan\beta$, $m_{0_{(3)}}-\tan\beta$,
$m_{0_{(1,2)}}-\mu$ and $m_{0_{(3)}}-\mu$ planes. Grey points satisfy REWSB and yield LSP neutralino.
Aqua points satisfy mass bounds including $1$ TeV $\lesssim m_{\tilde g}, m_{\tilde q} \lesssim 3.5$ TeV,
$m_{\tilde t_1} \gtrsim 0.7$ TeV,
$124~{\rm GeV}\leqslant m_h\leqslant 126~{\rm GeV}$ and B-physics bounds. Magenta points are subset of green points and also represent solutions with $\Omega_d h^2 \lesssim$ 1. Green points form a subset of magenta points and represent WMAP9 $3\sigma$ bounds { $0.1088\lesssim \Omega_d h^2 \lesssim 0.1217$}.
}
\label{funda1}
\end{figure}


\subsection{Results \label{spectroscopySection}}
In Figure~\ref{funda1} we display our results in $m_{0(1,2)}-M_{1/2}$, $m_{0(3)}-M_{1/2}$, $m_{0(1,2)}-\tan\beta$,
$m_{0(3)}-\tan\beta$, $m_{0(1,2)}-\mu$ and $m_{0(3)}-\mu$ planes. Grey points satisfy REWSB and neutralino
as an LSP conditions. Aqua points satisfy mass bounds including $1$ TeV
$\lesssim m_{\tilde g}, m_{\tilde q} \lesssim 3.5$ TeV, $m_{\tilde t_1} \gtrsim 0.7$ TeV,
$124~{\rm GeV}\leqslant m_h\leqslant 126~{\rm GeV}$, and B-physics bounds. Magenta points are subset of
green points and also represent solutions with $\Omega_d h^2 \lesssim$ 1. Green points form a subset of
magenta points and represent WMAP9 $3\sigma$ bounds $0.1088 \lesssim \Omega_d h^2 \lesssim 0.1217$. In these figures and the figures we
will be presenting below, there are patches of points. These patches represent scans with narrow parameter ranges
around phenomenologically interesting points. In the top left panel we see that magenta points are in the mass range
of $0.3 \,{\rm TeV} \lesssim m_{0(1,2)} \lesssim 2.5 \,{\rm TeV}$, while for $M_{1/2}$ mass range is 0.8 to 1.6 TeV.
When we insist on the stringent relic density bounds, we see two islands of green points within
$m_{0(1,2)} \lesssim 1 \,{\rm TeV}$ and $M_{1/2} \lesssim 1.7 \, {\rm TeV}$. It is important to
note that by generating more data the gap between these islands can be filled. Similarly in the top right panel we see
that magenta points are spread in the mass range of $0.5 \,{\rm TeV} \lesssim m_{0(1,2)} \lesssim 3 \,{\rm TeV}$.
From the middle left and right panels we note that in our scans, the minimum value of $\tan\beta \approx 9$ without
imposing relic density bounds but when do, then the range for $\tan\beta$ is 14-20. Plots in the bottom left and right
panels tell us that the parameter $\mu$ has a range of 0.6 to 3 TeV in our present data. But the data satisfying
relic density bounds is restricted to $2.5 \, {\rm TeV} \lesssim \mu \lesssim 3\,{\rm TeV}$. This means that these
green points do not consist of bino-higgsino mixed dark matter solutions and in our results we will not anticipate
solutions with large neutralino-nucleon scattering cross sections.

\begin{figure}
\centering
\subfiguretopcaptrue
\subfigure{
\includegraphics[width=6cm]{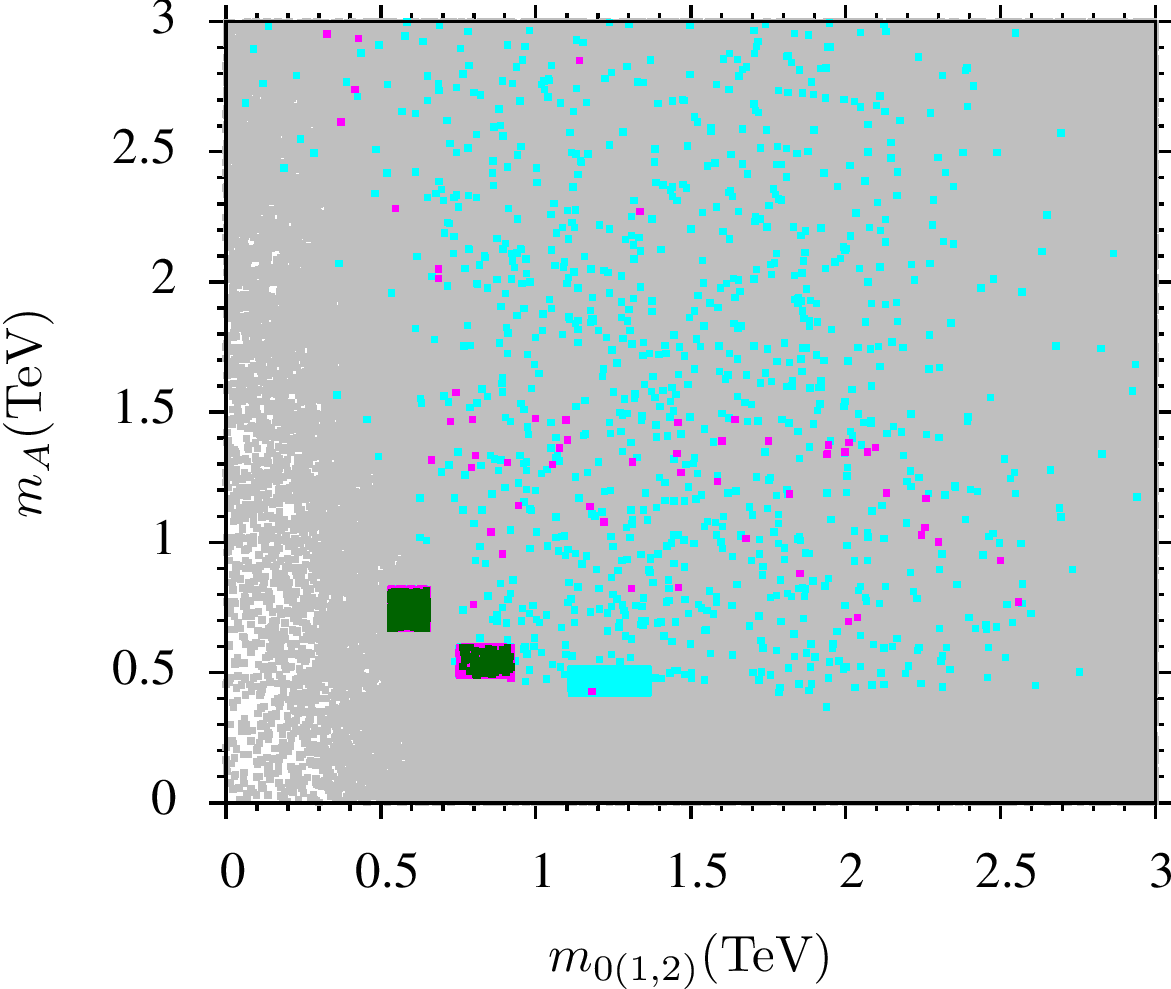}
}
\subfigure{
\includegraphics[width=6cm]{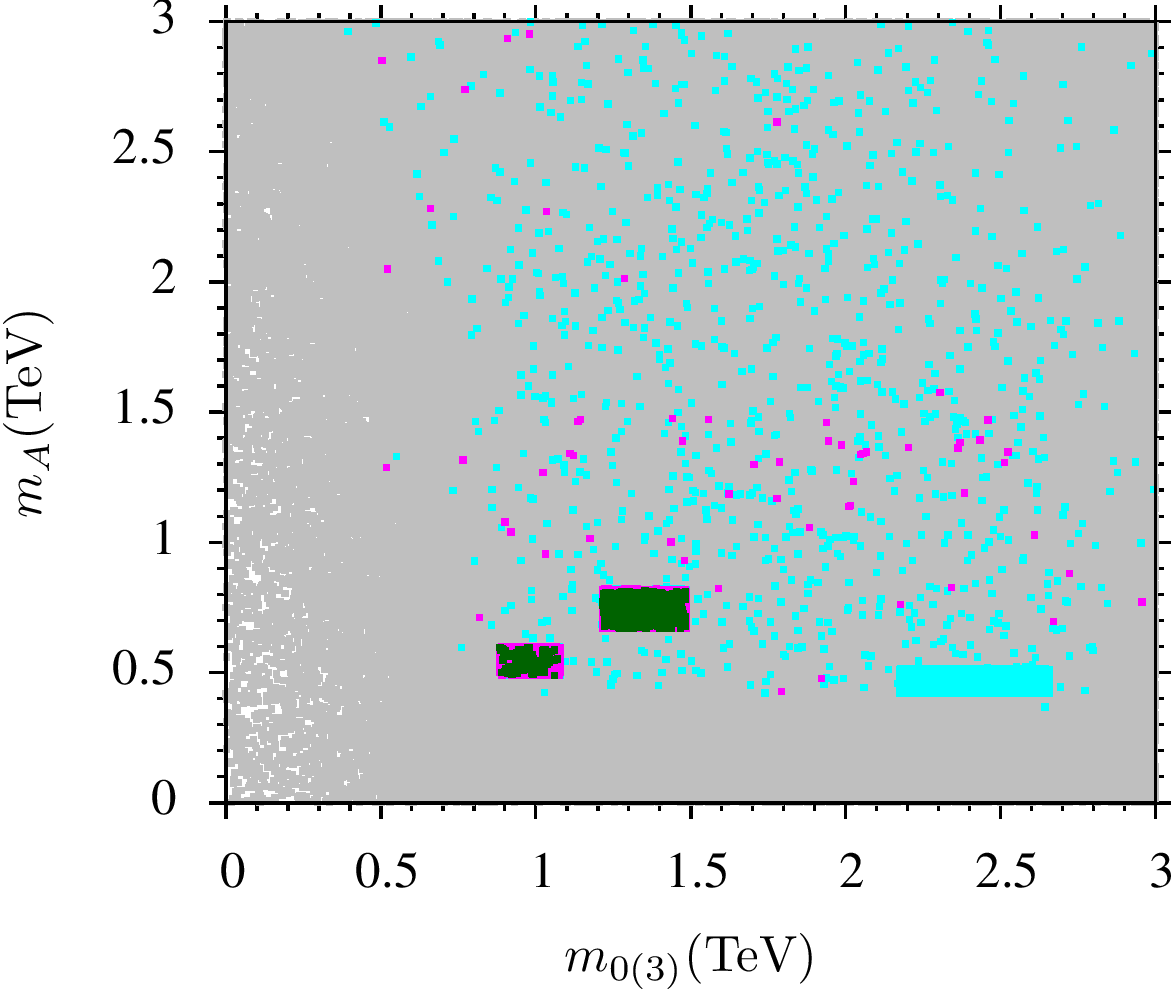}
}
\subfigure{
\includegraphics[width=6cm]{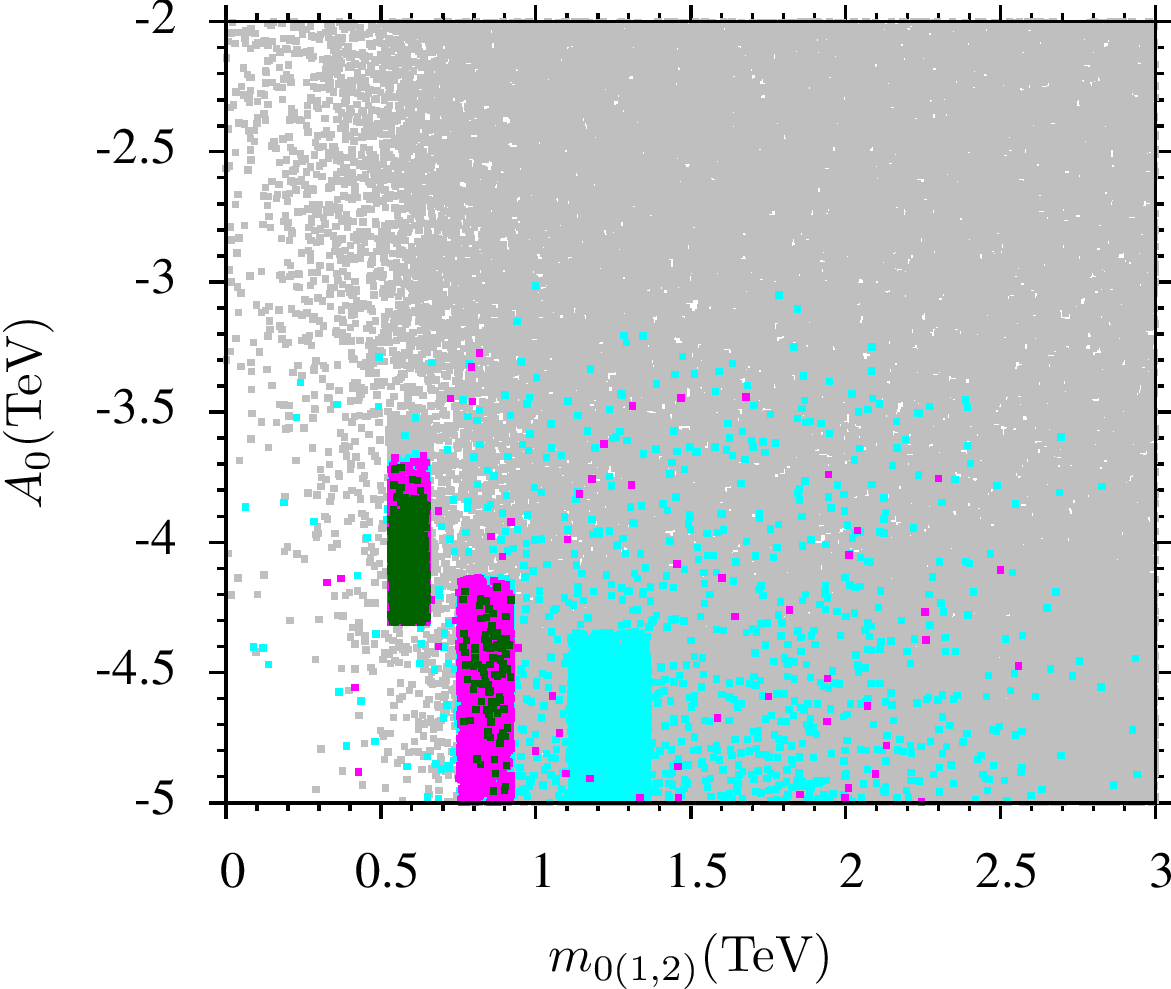}
}
\subfigure{
\includegraphics[width=6cm]{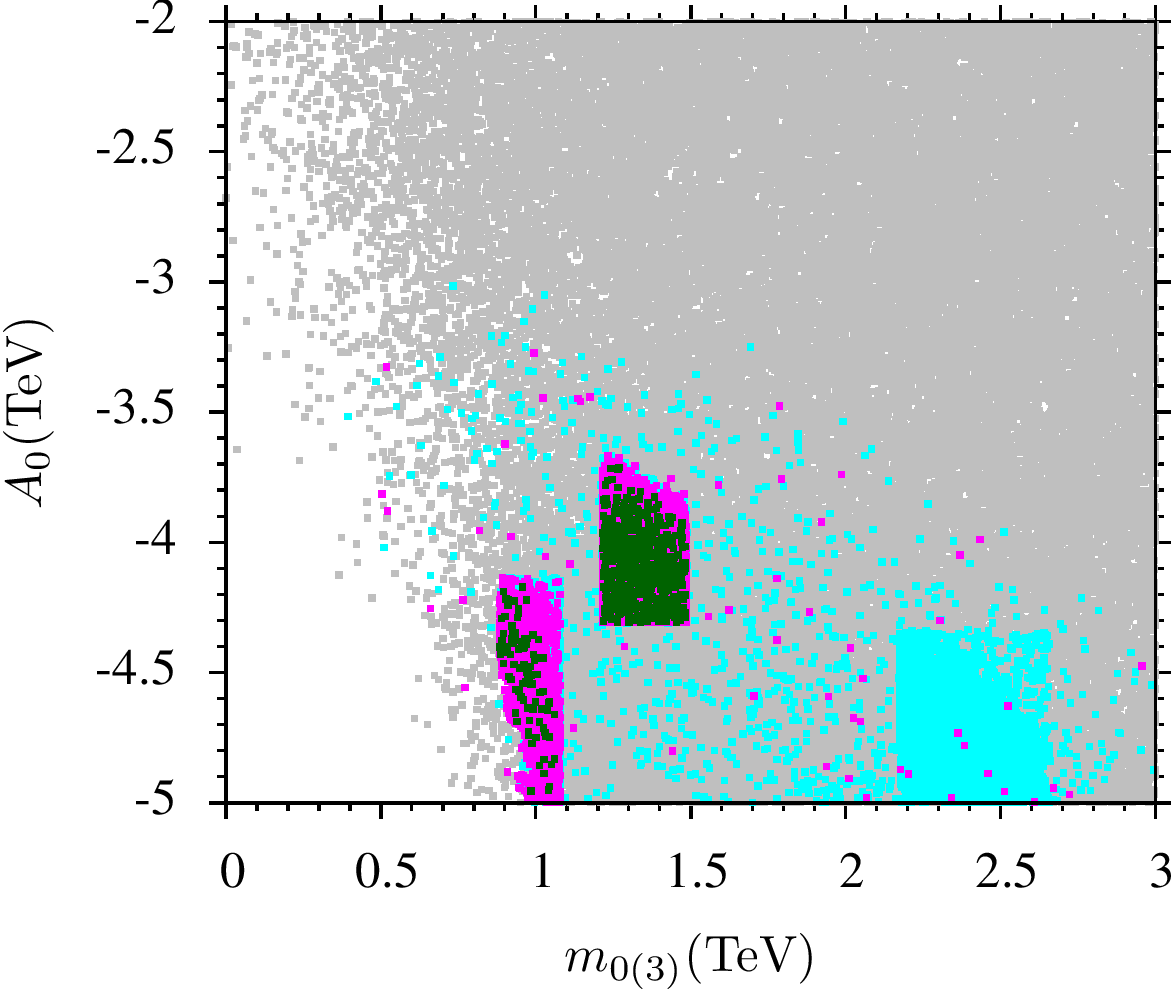}
}
\subfigure{
\includegraphics[width=6cm]{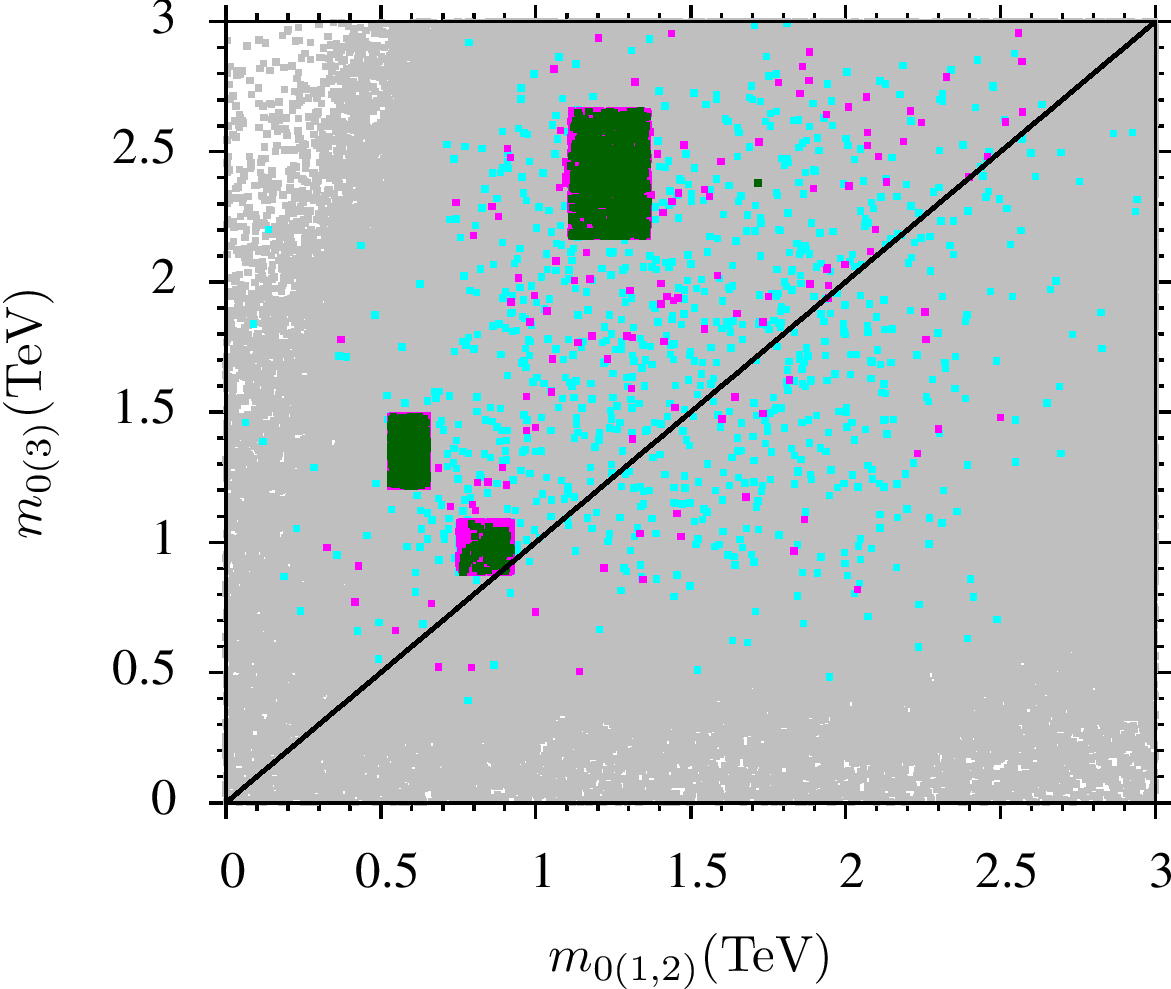}
}
\subfigure{
\includegraphics[width=6cm]{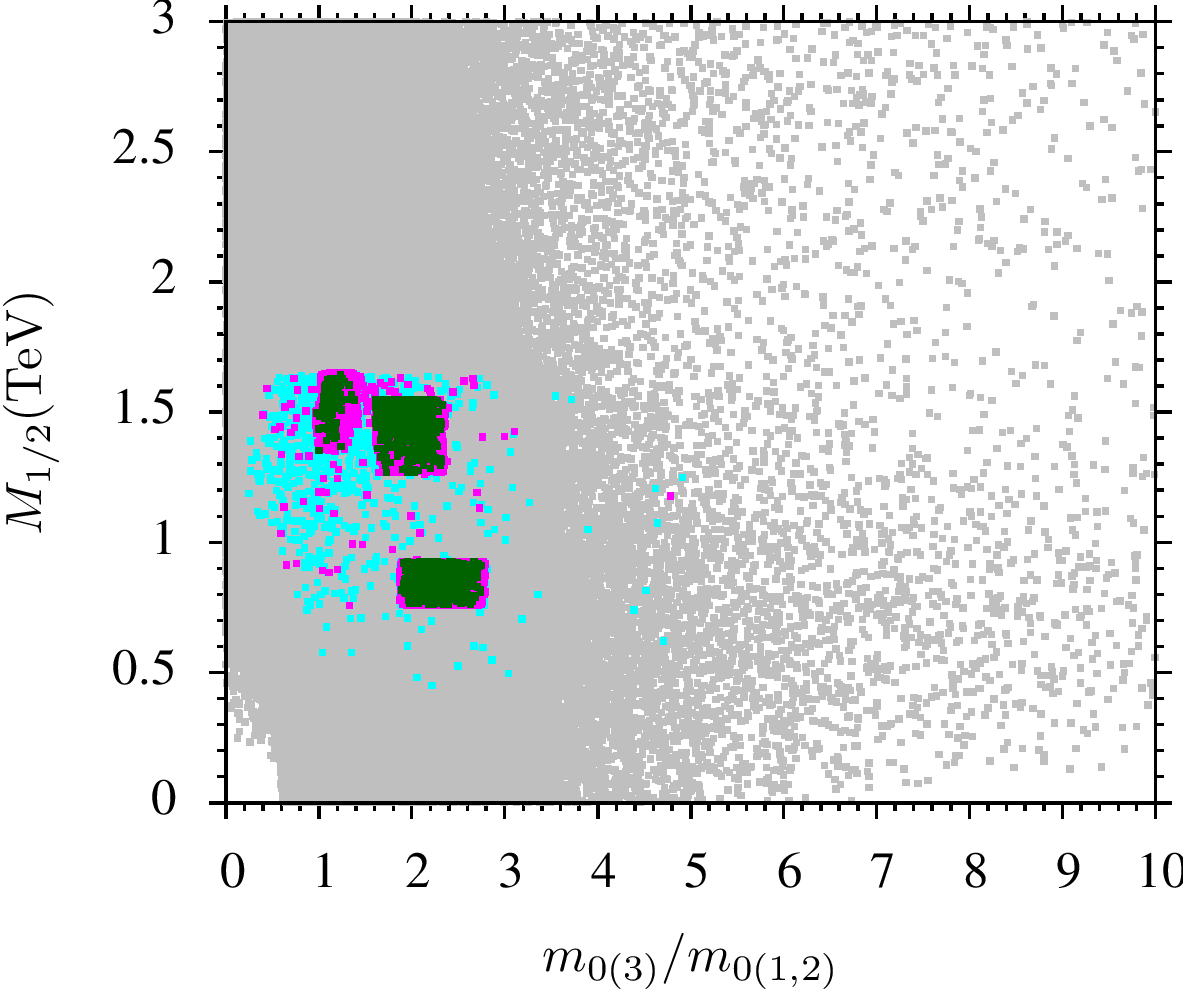}
}
\caption{Plots in $m_{0_{(1,2)}}-m_{A}$, $m_{0_{(3)}}-m_{A}$, $m_{0_{(1,2)}}-A_0$, $m_{0_{(3)}}-A_0$, $m_{0_{(1,2)}}-m_{0_{(3)}}$ and
$m_{0_{(3)}}/m_{0_{(1,2)}}-M_{1/2}$. Color coding same as in Figure~\ref{funda1}.
}
\label{funda2}
\end{figure}

In Figure~\ref{funda2} we present plots in $m_{0(1,2)}-m_{A}$, $m_{0(3)}-m_{A}$, $m_{0(1,2)}-A_0$,
$m_{0(3)}-A_0$, $m_{0(1,2)}-m_{0(3)}$ and $m_{0(1,2)}/m_{0(3)}-M_{1/2}$ planes. Color coding is the same as in
Figure~\ref{funda1}. From the top left and right
panels we note that the points (magenta points) consistent with particle mass bounds and B-physics bounds including bounds on Higgs mass given in Section~\ref{constraintsSection}, the  mass range for $m_A$ is 0.4 to 3 TeV but for the relic density consistent
points (green points), the mass range is 400 to 800 GeV. It is interesting to note that our results avoid
tough bounds set by BR($B_s \rightarrow \mu^{+}\mu^{-}$). We can understand this as follows. If we look at the plots containing
$\tan\beta$ as given in the middle panel of Figure~\ref{funda1}, we see that our solutions do not have large $\tan\beta$ values. Since in the MSSM $B_s \rightarrow \mu^{+}\mu^{-} \propto \tan\beta^{6}/m_A^{4}$, so the smallness of $\tan\beta$ with power 6
compensates the relatively small values of $m_A$. The middle left and right panels show that in our scans we
need $-5 \,{\rm TeV} \lesssim A_0 \lesssim -3 \, {\rm TeV}$ in case of magenta points and for green solutions
the range for $A_0$ is $-5$ to $-3.5$ TeV. In the bottom left panel we see that data consistent
with the constraints described in Section~\ref{constraintsSection}, the probable condition is $m_{0(1,2)} \lesssim m_{0(3)}$, though there are some points where $m_{0(1,2)} \approx m_{0(3)}$ and $m_{0(1,2)} \gtrsim m_{0(3)}$.

\begin{figure}
\centering
\subfiguretopcaptrue
\subfigure{
\includegraphics[width=6cm]{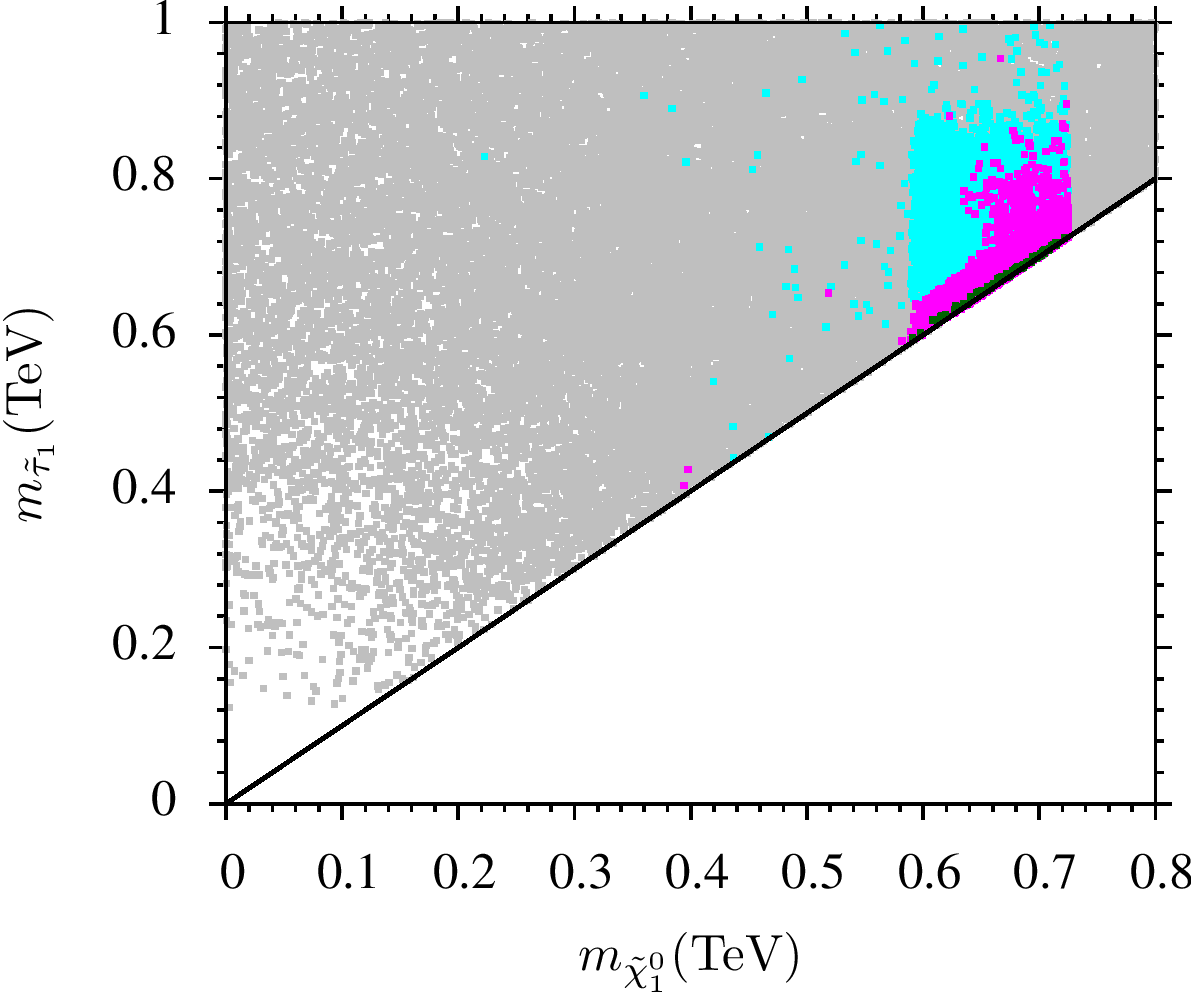}
}
\subfigure{
\includegraphics[width=6cm]{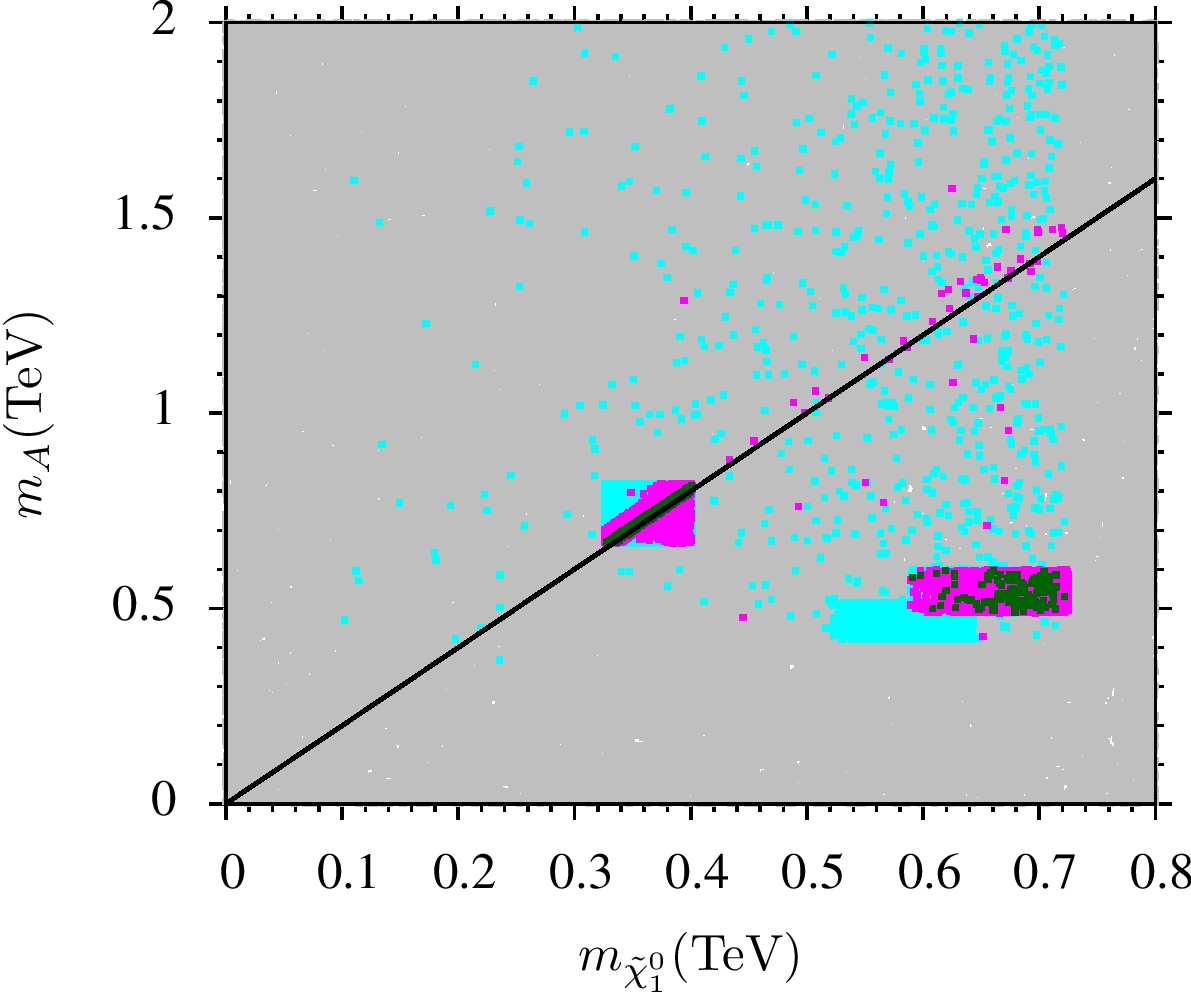}
}
\subfigure{
\includegraphics[width=6cm]{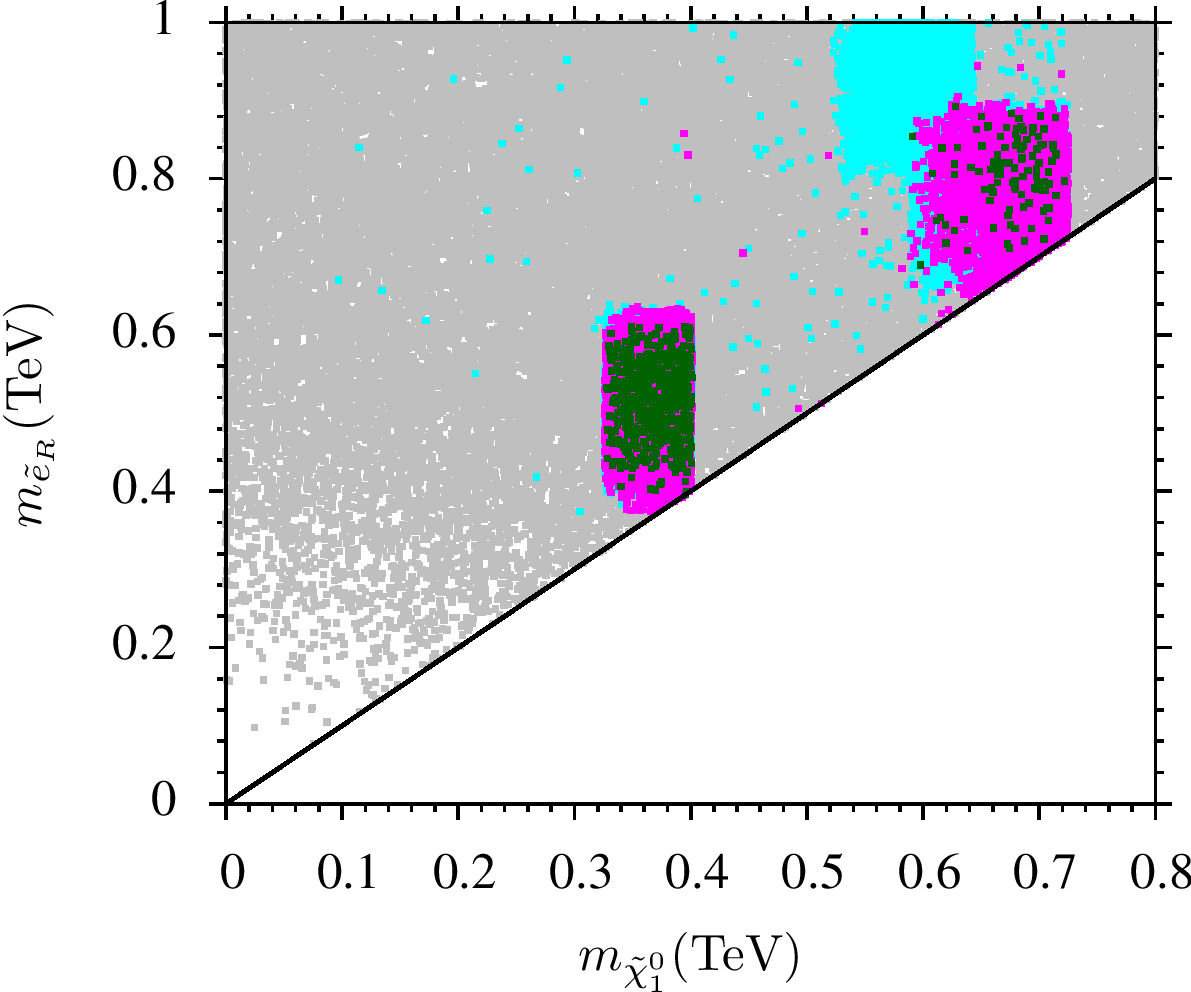}
}
\subfigure{
\includegraphics[width=6cm]{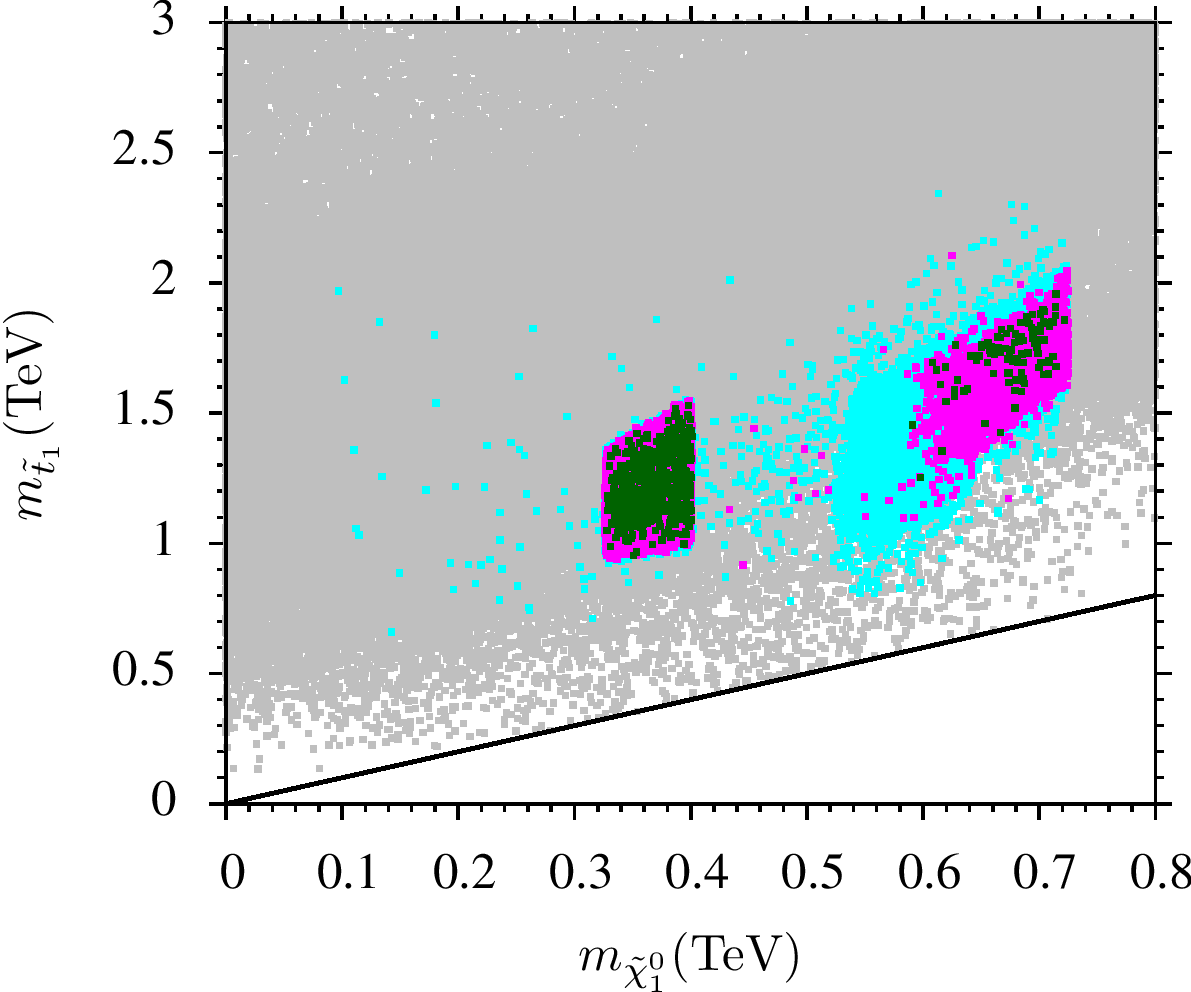}
}
\caption{Plots in $m_{\tilde \chi_{1}^{0}}-m_{\tilde \tau_1}$, $m_{\tilde \chi_{1}^{0}}-m_{A}$,
$m_{\tilde \chi_{1}^{0}}-m_{\tilde e_{R}}$ and $m_{\tilde \chi_{1}^{0}}-m_{\tilde t_1}$ planes.
 Color coding same as in Figure~\ref{funda1}.
Solid black lines are just to guide the eyes, where we can expect to have coannihilation and resonance solutions.
Note that in the neutralino-stop plane, the bound $m_{\tilde t_1}\gtrsim 0.7$ has not been applied.}
\label{spectrum}
\end{figure}


Plots in $m_{\tilde \chi_{1}^{0}}-m_{\tilde \tau}$, $m_{\tilde \chi_{1}^{0}}-m_{A}$,
$m_{\tilde \chi_{1}^{0}}-m_{\tilde e_{R}}$ and $m_{\tilde \chi_{1}^{0}}-m_{\tilde t_{1}}$ are shown in
Figure~\ref{spectrum}. Solid black lines are just to guide the eyes, where we can expect to have coannihilation and resonance solutions.
In the top left panel it is clearly visible that our model accommodate neutralino-stau coannihilation.
The NLSP stau mass and LSP neutralino mass degenerate in the mass range 400 to 740 GeV. For points strictly
consistent with relict density bounds, $m_{\tilde \tau_1}$ mass is  restricted to lie in the range 600 GeV to 740 GeV.
The right top panel shows our $A$-resonance solutions. It can be seen that our $A$-resonance solutions have minimum and
maximum values around $m_A \approx $ 0.7 TeV and 1.5 TeV respectively. The bottom left panel displays
neutralino-$\tilde e_R$ coannihilation solutions. Here we see two separate patches of points around $m_{\tilde e_R}\approx$ 400 GeV and 760 GeV. There are some points between the just mentioned mass ranges of $\tilde e_{R}$. A more exhaustive study
can fill up this gap. But it is clear enough to see the presence of neutralino-$\tilde e_R$ coannihilation scenario in our studies. In the bottom right panel we see that we do not have neutralino-stop coannihilation solutions. This is
because it usually requires large values of scalar masses ($\gtrsim $ 3 TeV). We can see in $m_{0_{(3)}}-A_{0}$ plane of
Figure~\ref{funda2} that even though for green points $|A_{0}|$ have relatively large values but the corresponding values for
 $m_{0_{(3)}}$ are below 1.5 TeV. It is because of it we are not having cancellation between diagonal and off-diagonal
terms in stop-mass matrix and we do not have light stop solutions.

\begin{figure}
\centering
\subfiguretopcaptrue
\subfigure{
\includegraphics[width=7cm]{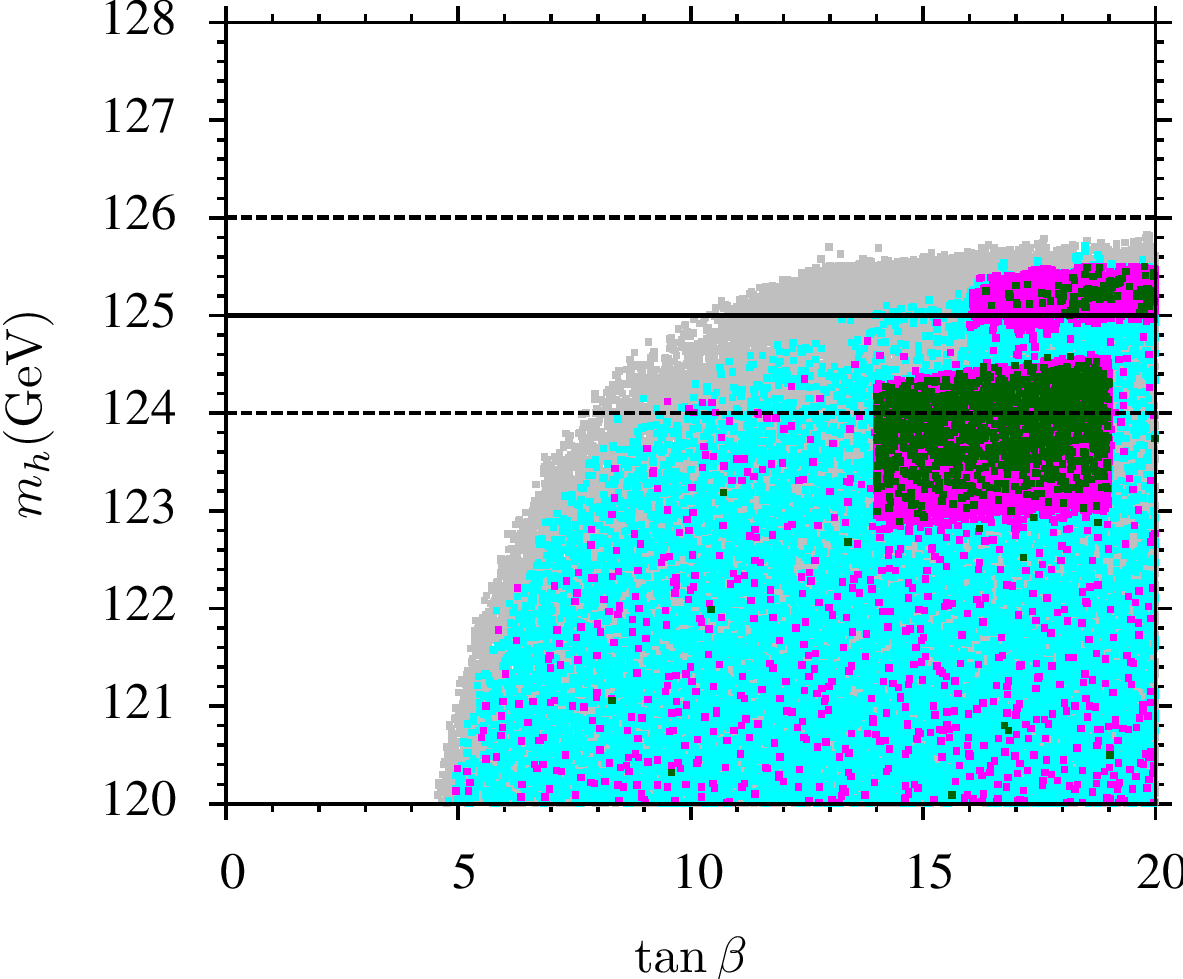}
}
\caption{Plot in $m_{h}-\tan\beta$ plane. Color coding same as in Figure~\ref{funda1}.
Solid and dashed horizontal lines represent 125 GeV, 123 GeV and 126 GeV respectively.}
\label{mhtanb}
\end{figure}
Plot in $\tan\beta-m_{h}$ plane is shown in Figure~\ref{mhtanb}. Color coding is the same as in Figure~\ref{funda1}.
Solid black line represents $m_h$ =125 GeV while dashed black lines represent bounds on Higss mass of 124 GeV
and 126 GeV respectively. We immediately see that in case of magenta points, we have
$m_h \approx$ 124 GeV for $\tan\beta \approx$ 9. On the other hand for data consistent with all the constraints given in
Section~\ref{constraintsSection}, we need $14 \lesssim \tan\beta \lesssim 20$.

\begin{figure}[t]
\centering
\subfiguretopcaptrue
\subfigure{
\includegraphics[width=7cm]{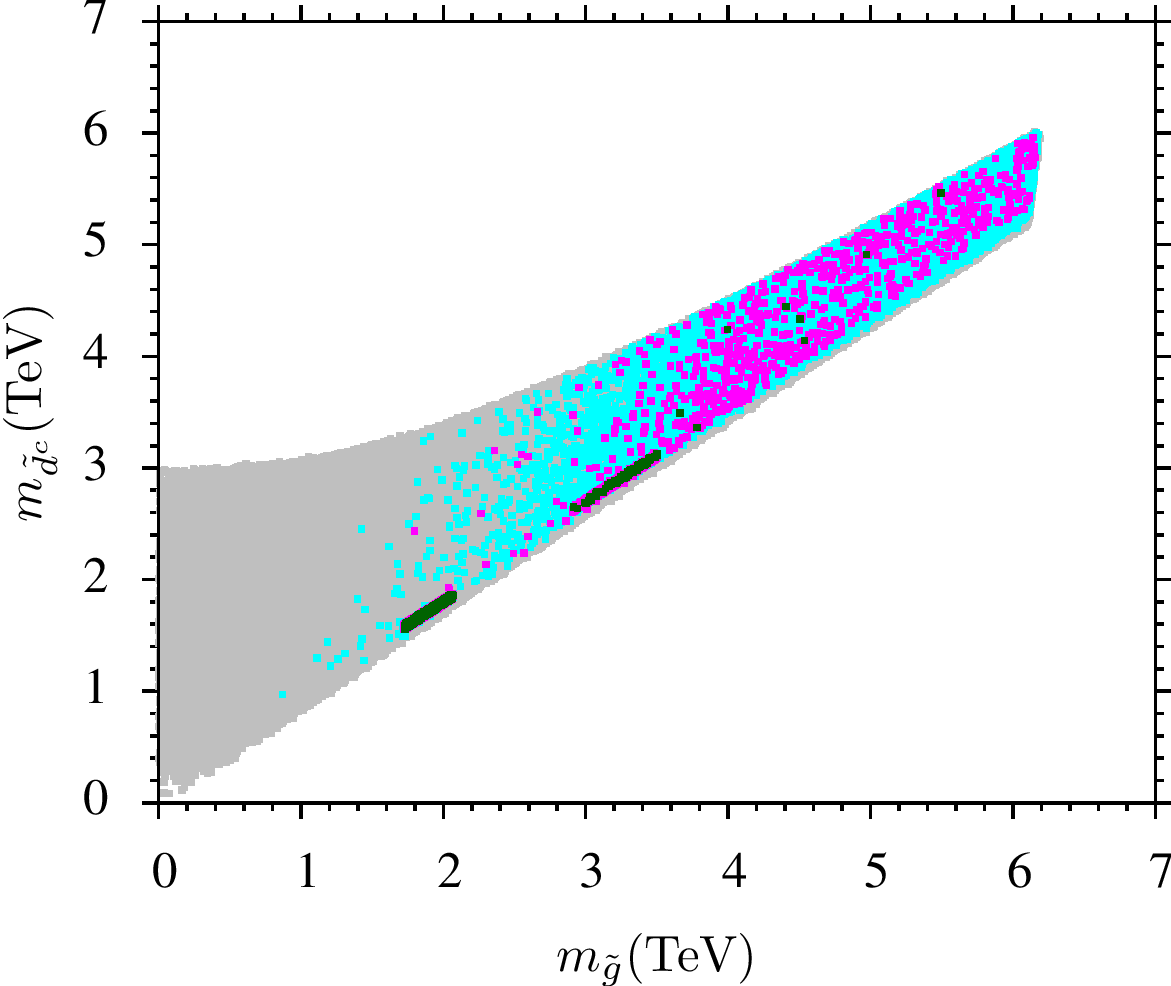}
}
\caption{Plot in $m_{\tilde g}-m_{\tilde d^{c}}$ plane. Color coding same as in Figure~\ref{funda1}.
}
\label{gludr}
\end{figure}

The LHC is a color particle producing machine. Gluino is considered to be the smoking gun of SUSY. In recent
LHC searches $m_{\tilde{g}} \gtrsim  1.5 \, {\rm TeV}\ ({\rm for}\ m_{\tilde{g}}\sim m_{\tilde{q}})
~\rm{and}~ m_{\tilde{g}}\gtrsim 0.9 \, {\rm TeV}\ ({\rm for}\ m_{\tilde{g}}\ll m_{\tilde{q}})$ \cite{Aad:2012fqa,Chatrchyan:2012jx} have been
obtained. We show results of our scans for gluino and squrak masses in Figure~\ref{gludr}.
It is evident from this figure that our solutions avoid the limits we just have mentioned. Here we see that
the minimum value for gluino mass is about 1.8 TeV for both green and magenta points, while for squark is about 1.6 TeV
in case of green points and 2.4 TeV for magenta points. We also see that the maximum value for
gluino mass is about 5.4 TeV with corresponding squark mass of about 5.6 TeV, while gluino and squarks masses may be as
heavy as 6.2 TeV if we do not insist on 2$\sigma$ WMAP9 bounds. Ref. \cite{cern_note1} shows that the squarks/gluinos of 2.5 TeV, 3 TeV and 6 TeV may be probed by the LHC14, High Luminosity (HL)LHC14 and High Energy (HE) LHC33 respectively. This clearly shows that our models have testable predictions.

\begin{figure}[h!]
\centering
\subfiguretopcaptrue
\subfigure{
\includegraphics[width=7cm]{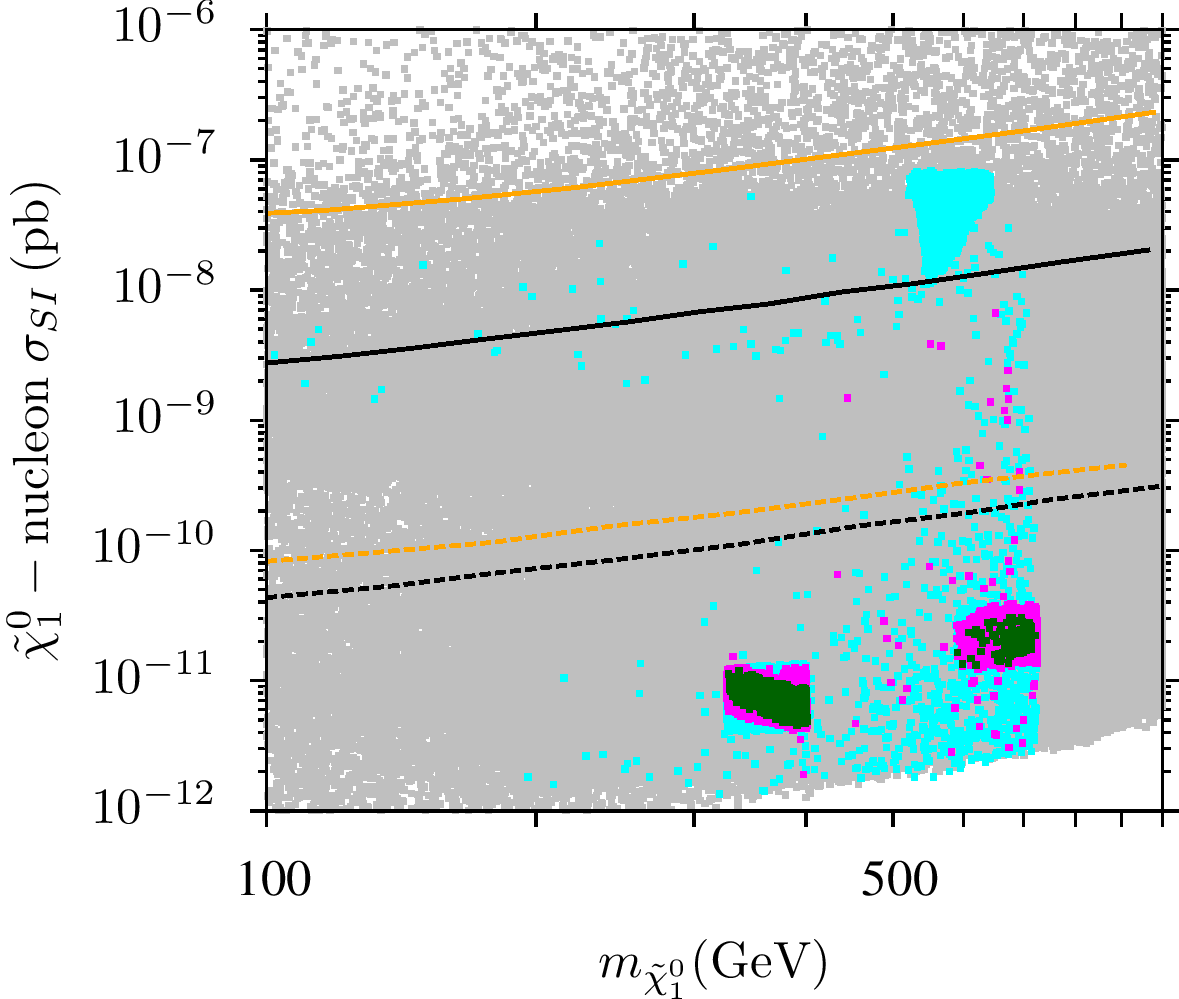}
}
\subfigure{
\includegraphics[width=7cm]{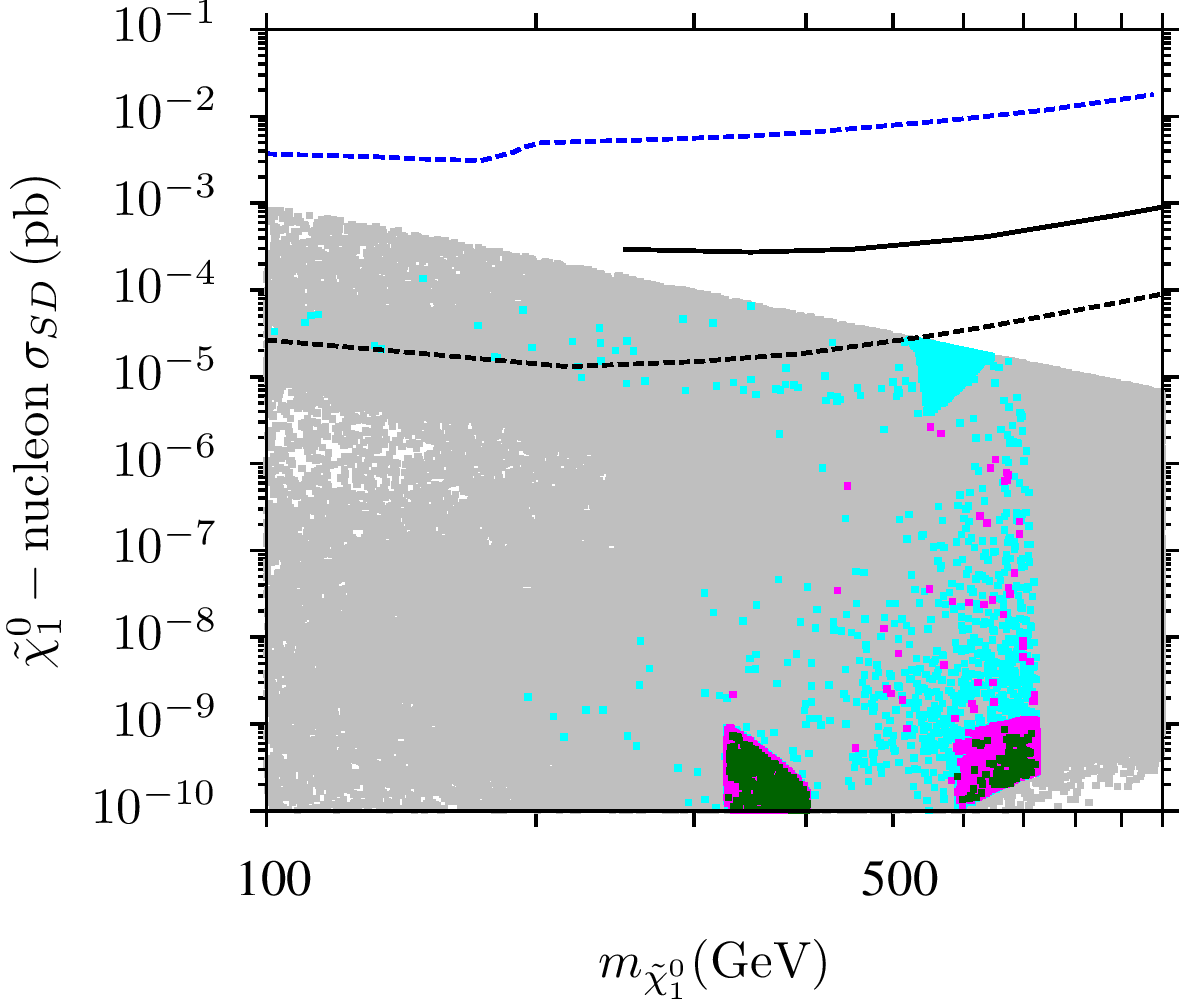}
}
\caption{Plots in $m_{\tilde \chi_{1}^{0}}-\sigma_{SI}$ and $m_{\tilde \chi_{1}^{0}}-\sigma_{SD}$ planes. Color coding same as in Figure~\ref{funda1}.
In the left panel, the orange solid line represents the current
 upper bound set by CDMS experiment, black solid line depicts upper bound set by XENON 100,
 while the orange (black) dashed line represents future reach
 of SuperCDMS (XENON 1T).
In the right panel, the current upper bounds set
 by Super-K (blue dashed line) and IceCube DeepCore (black solid line)
 are shown.
Future IceCube DeepCore bound is depicted by the black dashed line.
}
\label{Xsection}
\end{figure}

In Figure~\ref{Xsection} we show the spin independent and spin dependent cross sections as a function of the
neutralino LSP mass.
Color coding is the same as in Figure~\ref{funda1}. In the left panel, the orange solid line represents the current
 upper bound set by CDMS experiment, black solid line depicts upper bound set by XENON 100 \cite{Xenon100},  while the
orange (black) dashed line represents future reach  of SuperCDMS \cite{Brink:2005ej}(XENON 1T \cite{Xenon1T}). In the right panel, the current
upper bounds set by Super-K \cite{SuperK} (blue dashed line) and IceCube DeepCore (black solid line)  are shown. Future IceCube
DeepCore bound is depicted as the black dashed line \cite{IceCube}. Here we notice that green solutions are below the current
and future reaches of direct and indirect bounds. In the case of neutralino-nucleon spin-independent cross section
our solutions have cross section in the range of $10^{-12}$ to $10^{-11} \, {\rm pb}$ while for neutralino-nucleon
spin dependent case green points have cross section between $10^{-10}$ to $10^{-9} \, {\rm pb}$. Although our solutions have
very small cross sections which are hard to probe with the dark matter experiments mentioned above,  as stated earlier,
our models can be tested in collider searches.

In Table~\ref{table2}, we list three benchmark points. All of these points satisfy the sparticle mass and $B$-physics and relic density bounds as described in Section~\ref{constraintsSection}. Point 1 shows an  $A$-resonance solution. For this point, $m_A\approx$ 805 GeV,
$m_h$= 125 GeV, $m_{\tilde g}\approx$ 2058 GeV, first two family squarks are just under 2 TeV while sleptons of first two
families are less than 1 TeV, lighter stop is the lightest color sparticle. Points 2 displays a neutralino-$\tilde e_R$
coannihilation solution. In this example, $m_{\tilde e_{R}}\approx$ 413 GeV, $m_{h}=$ 124 GeV, gluino mass is $\sim$ 2040 GeV,
squarks are about 2 TeV, here too $\tilde t_1$ is the lightest colored particle, sleptons of third family are greater than
1 TeV but less than 1.6 TeV. An example of neutralino-stau coannihilation is shown in point 3, where we have
$m_{\tilde \tau_1}\approx$ 623 GeV, $m_h=$ 125 GeV, $m_{\tilde g}\sim  3041$ GeV, first two families squarks are under 3 TeV,
like points 1 and 2, here too, $\tilde t_1$ is the lightest colored sparticle, while the first two families sleptons are
relatively light.  Let us make a comment here. It can be seen from table~\ref{table2} that the input values of points 1 and 2
are almost similar. It occurs because we have neutralino-$\tilde e_R$ coannihilation scenario
(around $m_{\tilde \chi_{1}^{0}}\approx$ 400 GeV), where $m_{\tilde e_{R}}$ is also close to $m_A$ in mass, and
$m_A$ satisfies roughly the $A$-resonance condition ($m_{A} \approx 2m_{\tilde \chi_{1}^{0}}$) for points 1 and 2. We have a similar situation for $m_{\tilde \chi_{1}^{0}}\approx$ 700 GeV or so where $m_{\tilde e_{R}}$ is close in mass with
$m_{\tilde \tau_1}$. For example in point 3, $m_{\tilde \tau_1}\approx$ 623 GeV while $m_{\tilde e_{R}}\approx$ 718 GeV. Had
we chosen a point from such a region, we would have had point 2 and point 3 similar.


\begin{table}[h!]
\centering
\scalebox{1.0}{
\begin{tabular}{lccc}
\hline
\hline
                 & Point 1 & Point 2 & Point 3       \\
\hline
$M_{1/2}$          & 922.4   & 915.1   & 1419      \\
$m_{0(1,2)}$       & 621.1   &550.8    & 770.9         \\
$m_{0(3)}$         & 1256    &1437     & 895.6        \\
$\tan\beta$        &  16.98  & 15.45   & 18.15       \\
$A_{0}$            & -4255  & -4246    & -4189            \\
$\mu$              &  3112  & 2656     &  2662           \\
$m_{A}$            & 804.8   & 798.6   & 545.8            \\
$sign(\mu)$        & +      & +   & +        \\
\hline
$m_h$            &125  & 124  & 125    \\
$m_H$            & 809 & 803 & 549        \\
$m_{H^{\pm}}$    & 814 & 808 & 555       \\

\hline
$m_{\tilde{\chi}^0_{1,2}}$
                 &400, 763   & 396, 755      &620, 1168         \\
$m_{\tilde{\chi}^0_{3,4}}$
                 &3087, 3087  &2635, 2636    &2644,2645     \\

$m_{\tilde{\chi}^{\pm}_{1,2}}$
                 &766, 3088   &758, 2639   &1172, 2649    \\
$m_{\tilde{g}}$  & 2058      & 2040     & 3041    \\

\hline $m_{ \tilde{u}_{L,R}}$
                 & 1946, 1912   & 1906, 1893  & 2853, 2796     \\
$m_{\tilde{t}_{1,2}}$
                 & 1358, 1910  & 1141, 1872   & 1545, 2366      \\
\hline
$m_{ \tilde{d}_{L,R}}$
                 & 1948, 1860  & 1908, 1817  & 2854, 2717     \\
$m_{\tilde{b}_{1,2}}$ & 1863,2089  & 1843, 2179   & 2338, 2640     \\
\hline
$m_{\tilde{\nu}_{1}}$
                    & 918   &  889      &  1282        \\

$m_{\tilde{\nu}_{3}}$
                 &  1385     &  1560     &  1298        \\
\hline
$m_{ \tilde{e}_{L,R}}$
                &925 , 585   & 897, 413 & 1288,718     \\
$m_{\tilde{\tau}_{1,2}}$
                & 1122, 1388    & 1291, 1563  & 623, 1298     \\
\hline

$\sigma_{SI}({\rm pb})$
                &4.54 $\times 10^{-12}$ & 6.76$\times 10^{-12}$ & 2.31$\times 10^{-11}$ \\
$\sigma_{SD}({\rm pb})$
                &2.45 $\times 10^{-9}$ & 1.25$\times 10^{-9}$ & 2.26$\times 10^{-10}$ \\

$\Omega_{CDM}h^2$
                & 0.11  & 0.108    &  0.121     \\

\hline
\hline
\end{tabular}
}
\caption{ Three benchmark points for sMSSM. All masses in this table are in units of GeV.
All points satisfy the sparticle mass, $B$-physics and relic density constraints as described in Section~\ref{constraintsSection}.
Point 1 shows an $A$-resonance solution. Point 2 displays a solution with neutralino-$\tilde e_{R}$ nearly degenerate in mass leading
to coannihilation. Point 3 is an example of neutralino-stau coannihilation solution.
}
\label{table2}
\end{table}

\section{ Conclusion \label{conclusions}}

In this paper we have proposed and developed a class of supersymmetric models in which the SUSY breaking Lagrangian takes
the most general form consistent with two symmetry requirements.  The first is compatibility with a grand unified symmetry
such as $SO(10)$, and the second is consistency with a non-Abelian flavor symmetry $H$.  This symmetry is chosen so that
flavor violation arising from SUSY particle exchange is sufficiently suppressed.  A minimal class of models, termed
sMSSM -- for flavor symmetry-based minimal supersymmetric standard model -- is constructed.  We also investigated the
phenomenology of sMSSM in some detail.

We have constructed four explicit models which generate a common spectrum of SUSY particles at low energies.  The first
two are based on $SU(2)_H$ and $SO(3)_H$ local gauge symmetries with a ${\bf 2+1}$ and {\bf 3} family assignment respectively.
In both cases we have found simple solutions to the $D$-term problem that can potentially induce large flavor violation.
An interchange symmetry acting on the doublets of $SU(2)_H$ model sets the $D$-term to zero.  In the $SO(3)_H$ model,
{ an} explicit potential is constructed that makes the $D$-term { vanish} by virtue of a $Z_2 \times Z_2$ symmetry acting on
the scalar fields.  We also present two models based on discrete non-Abelian symmetries $S_3$ and $A_4$ with either
a ${\bf 2+1}$ or a ${\bf 3}$ family assignment.  These models, being discrete, do not have any issue with $D$-terms.
We have analyzed in all models the symmetry breaking sector and ensured that realistic fermion masses are generated
without inducing excessive SUSY flavor violation.

Our phenomenological analysis uses as input the seven parameters listed in Eq. (\ref{para}).  This is slightly larger
than the four parameter set of cMSSM, but still small enough to carry out a detailed parameter scan.  Our results show
that almost all SUSY particles can have masses below about 3 TeV, consistent with radiative electroweak symmetry
breaking and $B$-physics constraints.  We also find solutions with the correct relic density of neutralino dark matter,
although their direct detection would be difficult in dark matter experiments.  When the LHC resumes operation,
we are optimistic that it would discover supersymmetric particles with properties predicted by sMSSM. We have
presented three benchmark points in Table \ref{table2}, which also shows that the fine tuning needed in these models
is relatively mild.


\section*{Acknowledgments}
This work is supported in part by the DOE Grant No. de-sc0010108 (KSB) and by DOE grant No. DE-FG02-12ER41808 (IG and QS).
This work used the Extreme Science and Engineering Discovery Environment (XSEDE), which is
supported by the National Science Foundation grant number OCI-1053575. IG acknowledges support from the  Rustaveli
National Science Foundation  No. 31/98.
KSB and IG would like to thank CETUP* (Center for Theoretical Underground Physics and Related
Areas), supported by the US Department of Energy under Grant No. de-sc0010137 and by the
US National Science Foundation under Grant No. PHY-1342611, for its hospitality and partial
support during the 2013 Summer Program.  They also wish to thank Barbara Szczerbinska for providing
a stimulating atmosphere in Deadwood during the CETUP* 2013 program.


\end{document}